\documentclass[aps,preprintnumbers,10pt,superscriptaddress,twocolumn]{revtex4-1}

\usepackage[dvips]{graphicx}

\usepackage{siunitx}
\usepackage{array,amsmath,amsthm,feynmf,mathtools}
\usepackage{multirow}
\usepackage{mathtools}
\usepackage{epsfig}
\usepackage{graphicx}
\usepackage{xcolor}
\usepackage{slashed}
\usepackage{amssymb}
\usepackage{bm}
\usepackage{cancel}
\usepackage[normalem]{ulem}
\usepackage[export]{adjustbox}
\usepackage{enumitem}
\usepackage[colorlinks=true,linkcolor=blue,citecolor=blue]{hyperref}

\newcommand{\cw}[0]{c_\textsc{w}}
\newcommand{\sw}[0]{s_\textsc{w}}
\newcommand{\Qem}[0]{Q_\textsc{em}}
\newcommand{\Qw}[0]{Q_\textsc{w}}
\newcommand{\Li}[0]{\text{Li}_2}
\newcommand{\LiOM}[2]{\text{Li}_2\Big(1-\frac{#1}{#2}\Big)}

\newcommand{\sPhi}[2]{\Phi\big({\textstyle\frac{#1}{#2}}\big)}

\newcommand{\mHp}{m_{H^{\smash{+}}}}

\newcommand{\tHooft}[0]{\mbox{'\lowercase{t} Hooft }}

\begin{document}

\title{Electron EDM in the complex two-Higgs doublet model}
\author{Wolfgang Altmannshofer}
\email{waltmann@ucsc.edu}
\affiliation{Department of Physics and Santa Cruz Institute for Particle Physics\\ University of California, Santa Cruz, CA 95064, USA}

\author{Stefania Gori}
\email{sgori@ucsc.edu}
\affiliation{Department of Physics and Santa Cruz Institute for Particle Physics\\ University of California, Santa Cruz, CA 95064, USA}

\author{Nick Hamer}
\email{nhamer@ucsc.edu}
\affiliation{Department of Physics and Santa Cruz Institute for Particle Physics\\ University of California, Santa Cruz, CA 95064, USA}

\author{Hiren H. Patel}
\email{hpatel6@ucsc.edu}
\affiliation{Department of Physics and Santa Cruz Institute for Particle Physics\\ University of California, Santa Cruz, CA 95064, USA}

\begin{abstract}
We present the first complete two loop calculation of the electron EDM in the complex two-Higgs doublet model.  We confirm gauge-independence by demonstrating analytic cancellation of the gauge parameter $\xi$ in the background field gauge and the \tHooft $R_\xi$ gauge.  We also investigate the behavior of the electron EDM near the decoupling limit, and determine the short- and long-distance contributions by matching onto an effective field theory.  Compared with earlier studies of the electron EDM in the complex two-Higgs doublet model, we note disagreements in several places and provide diagnoses where possible.  We also provide expressions for EDMs of light quarks.
\end{abstract}

\maketitle

\section{Introduction}
The discovery of a non-vanishing electric dipole moment (EDM) of any fundamental particle in next generation experiments would unambiguously signal the existence of new sources of CP-violation beyond the Standard Model (SM) of particle physics.  Indeed, many such models predict EDMs of elementary particles that are within reach of current experiments, with the SM contribution estimated to lie several orders of magnitude lower~\cite{Pospelov:2013sca,Yamaguchi:2020eub,Yamaguchi:2020dsy}.  Such a discovery could supply a crucial ingredient towards solving the long standing problem of the origin of the cosmic baryon asymmetry \cite{Pospelov:2005pr,Engel:2013lsa}.  Currently, the most stringent limit on the electron EDM is provided by the ACME collaboration \cite{Andreev:2018ayy} and reads $d_e < 1.1\times 10^{-29} e\text{ cm}$ at a 90\% confidence level.  The collaboration expects an improvement in sensitivity by an order of magnitude in the near future \cite{Andreev:2018ayy}.  A further significant improvement in sensitivity might come in the future from the EDM$^3$ experiment \cite{Vutha:2018tsz}. 

Two-Higgs doublet models (2HDMs) are among the most popular extensions of the SM and can contain new sources of CP-violation. 2HDMs arise in many well-motivated theories beyond the SM, such as in the Minimal Supersymmetric Standard Model (MSSM). The most general form of a 2HDM allows for new sources of CP-violation both in the scalar potential and in the Higgs-Yukawa interactions. However, it generically exhibits flavor changing neutral currents, which are strongly constrained by experiments.  By imposing a softly broken $\mathbb{Z}_2$ symmetry \cite{Glashow:1976nt} to yield the complex two-Higgs doublet model (C2HDM), flavor changing neutral currents at tree-level are naturally eliminated. The $\mathbb{Z}_2$ symmetric C2HDM still accommodates new sources of CP-violation in the scalar potential to generate EDMs of fundamental particles.

Analyses of electric dipole moments in the C2HDM have a long history, starting with the calculation of two loop Barr-Zee diagrams \cite{Barr:1990vd}, followed by several extensions, e.g. \cite{Leigh:1990kf,Gunion:1990ce,Chang:1990sf}. However, the results of these previous works only include a subset of all two loop contributions and are not gauge-invariant. More recently, Ref. \cite{Abe:2013qla} employed the pinch technique to calculate the Barr-Zee diagrams gauge invariantly.  Still, as indicated by the authors, not all contributions to the electron EDM were included.

In this paper, we present for the first time the complete calculation of the electron EDM by systematically accounting for all Feynman diagrams that contribute at two loop order.  Due to the recurrent issue of gauge-invariance, we perform the calculation in both the background field gauge \emph{and} in the conventional \tHooft $R_\xi$ gauge keeping the gauge parameter $\xi$ arbitrary.  We algebraically establish $\xi$-independence and reach agreement in both gauges providing strong validation for our results.  Our final formula for the electron EDM in the C2HDM is given in (\ref{eq:gaugeinv-totaledm}).  This is the main equation that should be used in phenomenological exploration of the electron EDM.  For convenience, we provide a \emph{Mathematica} notebook containing the necessary formulae as an ancillary file. 

The presentation of our work is organized as follows.  In Sec.~\ref{Sec:C2HDM}, we introduce the C2HDM, establishing the notation we use in this paper.
In Sec.~\ref{Sec:BGF-eval}, we present the electron EDM in background field gauge. Our main results are contained in this section.  In Sec.~\ref{Sec:FeynmanGauge}, we reevaluate the EDM in the conventional Feynman-\tHooft gauge and explain how we reach agreement with the background field evaluation.  In Sec.~\ref{Sec:comparison}, we compare our results with the recent evaluation of the electron EDM presented in \cite{Abe:2013qla}.  We also introduce a set of benchmark parameters to carry out a numerical exploration of the electron EDM.  In Sec.~\ref{Sec:Lightquark}, we explain how our results may be adapted to obtain EDMs of light quarks.  In Sec.~\ref{Sec:Decoupling}, we present an asymptotic expansion of the electron EDM near the decoupling limit and discuss its relationship to the formula derived from an effective field theory.  Sec.~\ref{Sec:summary} is reserved for our conclusions.  Finally, in the appendix, we collect useful equations on the 2HDM scalar potential.
\newpage
\section{Formulation of the C2HDM}\label{Sec:C2HDM}
The C2HDM is the most general CP-violating two-Higgs doublet model that possesses a softly-broken $\mathbb{Z}_2$ symmetry.
In our discussion, we will closely follow the notation of \cite{Haber:2006ue,Boto:2020wyf}, to which we refer the reader for a detailed description of its formulation. 

The SM scalar sector is extended by an additional scalar doublet with identical quantum numbers as the SM Higgs.  The scalar potential is
\begin{multline}\label{eq:genpot}
V(\Phi_1,\,\Phi_2)=m_{11}^2\Phi_1^\dagger\Phi_{1} +
m_{22}^2\Phi_2^\dagger\Phi_{2} -
\Big(m_{12}^2\Phi_1^\dagger\Phi_{2}+\text{c.c.}\Big) \\
+ {\textstyle\frac{1}{2}}\lambda_1\big(\Phi_1^\dagger\Phi_{1}\big)^2
+{\textstyle\frac{1}{2}}\lambda_2\big(\Phi_2^\dagger\Phi_{2}\big)^2
+\lambda_3\big(\Phi_1^\dagger\Phi_{1}\big)\big(\Phi_2^\dagger\Phi_2\big) \\
+\lambda_4\big(\Phi_1^\dagger\Phi_{2}\big)\big(\Phi_2^\dagger\Phi_1\big)
 +\Big({\textstyle\frac{1}{2}}\lambda_5\big(\Phi_1^\dagger\Phi_2\big)^2+\text{c.c.}\Big)\,.
\end{multline}
Apart from the soft-breaking term proportional to $m_{12}^2$, the potential exhibits invariance under the $\mathbb{Z}_2$ transformation $\Phi_2 \rightarrow - \Phi_2$.  Generally, both doublets may acquire a vacuum expectation value.  Assuming the parameters are chosen to respect $\text{U}(1)_\text{EM}$ in the vacuum, they take the form
\begin{equation}
\langle\Phi_1\rangle = \frac{1}{\sqrt{2}}\begin{pmatrix}0\\v_1\end{pmatrix},\qquad
\langle\Phi_2\rangle = \frac{1}{\sqrt{2}}\begin{pmatrix}0\\v_2 e^{i\zeta}\end{pmatrix}\,,
\end{equation}
where $v\equiv \sqrt{v_1^2 + v_2^2} = 246\text{ GeV}$, and $\zeta$ is a possible relative phase between them.  The values of $v_1$, $v_2$, and $\zeta$ are given in terms of the potential parameters in the appendix.  We use rephasing invariance to work in the basis where $\zeta=0$ throughout the paper.
It is convenient to transform to the Higgs basis
\begin{equation}
\begin{pmatrix}
\Phi_1 \\ \Phi_2
\end{pmatrix}=
\begin{pmatrix}
\cos\beta & -\sin\beta\\
\sin\beta & \cos\beta
\end{pmatrix}
\begin{pmatrix}
H_1\\
H_2
\end{pmatrix}\,,
\end{equation}
with $\tan\beta = v_2/v_1$ so that the vacuum expectation value is contained entirely in $H_1$. In this new basis the potential reads
\begin{multline}\label{eq:HBpot}
\mathcal{V}(H_1,\,H_2)=
Y_1 H_1^\dagger H_{1} +
Y_2 H_2^\dagger H_{2} +
\Big(Y_3  H_1^\dagger H_{2} +\text{c.c.}\Big) \\
+{\textstyle\frac{1}{2}}Z_1\big( H_1^\dagger H_{1}\big)^2
+{\textstyle\frac{1}{2}}Z_2\big( H_2^\dagger H_{2}\big)^2 +Z_3\big( H_1^\dagger H_{1}\big)\big( H_2^\dagger H_2\big)\\
+Z_4\big( H_1^\dagger H_{2}\big)\big( H_2^\dagger H_1\big)
 + \Big({\textstyle\frac{1}{2}}Z_5  \big( H_1^\dagger H_2\big)^2 \\
 +\big(Z_6 \, H_1^\dagger H_1 + Z_7\, H_2^\dagger H_2 \big) H_1^\dag H_2+\text{c.c.}\Big)\mathclap{\,,}
\end{multline}
where the new parameters $Y_i$, $Z_i$ are linear combinations of the original parameters $m_{ij}^2$, $\lambda_i$ given the appendix.  Analysis of small fluctuations around the vacuum shows that the components of the scalar fields in the Higgs basis are given by
\begin{equation*}
H_1 =
\begin{pmatrix}
G^+\\
\frac{1}{\sqrt{2}}\big(v + \varphi_1^0 + i G^0\big)
\end{pmatrix}\enspace
H_2 =
\begin{pmatrix}
H^+ \\
\frac{1}{\sqrt{2}}\big(\varphi_2^0 + i a^0\big)
\end{pmatrix}\,,
\end{equation*}
where $G^+$, $G^0$ are the would-be Goldstone modes supplying the longitudinal modes of the massive $W$, $Z$ gauge bosons, and $H^+$ is a physical charged Higgs, of mass squared $\mHp^2 = Y_2 + \frac{1}{2}Z_3 v^2$.  The remaining scalars---the CP-even $\varphi_1^0$ and $\varphi_2^0$, and CP-odd $a^0$---mix, with the Higgs squared-mass matrix $\mathcal{M}^2$ given by
\begin{equation}\label{eq:neutralMassMtx}
 \frac{\mathcal{M}^2}{v^2}  = \left(
\begin{matrix}
Z_1 & \text{Re}(Z_6) & -\text{Im}(Z_6) \\[0.5mm]
    & Y_2/v^2+\frac{1}{2}Z^+_{345} & -\frac{1}{2}\text{Im}(Z_5 ) \\[1mm]
    & &  Y_2/v^2+\frac{1}{2}Z^-_{345}
\end{matrix}\right),
\end{equation}
where $Z_{345}^\pm = Z_3 + Z_4 \pm \text{Re}(Z_5) $.  The mass matrix is diagonalized by a special orthogonal matrix $R$
\begin{align}
\label{eq:DiagMassMtx}
R \mathcal{M}^2 R^\top ={}& \text{diag}(m_1^2,m_2^2,m_3^2)\\
\label{eq:HBtoMassRot}\begin{pmatrix}
h_1 \\ h_2 \\ h_3
\end{pmatrix} ={}& R
\begin{pmatrix}
\varphi_1^0 \\
\varphi_2^0 \\
a^0
\end{pmatrix}\,,
\end{align}
where we parameterize the elements of $R$ as
\begin{equation}\label{eq:qmtx}
R = \begin{pmatrix}
q_{11} & \text{Re}(q_{12}) & \text{Im}(q_{12})\\
q_{21} & \text{Re}(q_{22}) & \text{Im}(q_{22})\\
q_{31} & \text{Re}(q_{32}) & \text{Im}(q_{32})
\end{pmatrix}\,.
\end{equation}
Elements of $R$ are subject to orthonormality conditions
\begin{gather}
\label{eq:qnorm}\sum_{k=1}^3 q_{k1}^2 = \frac{1}{2}\sum_{k=1}^3 |q_{k2}|^2 = 1\,,\\
\label{eq:qorth}\sum_{k=1}^3 q_{k2}^2 = \sum_{k=1}^3 q_{k1} q_{k2} = 0\,,
\end{gather}
which prove indispensable in the calculation of the electron EDM.  Inserting the linear combinations (\ref{eq:HBtoMassRot}) into the scalar potential (\ref{eq:HBpot}) generates the interaction vertices in terms of mass eigenstate fields, for which we point the reader to \cite{Haber:2006ue} for a complete listing.  For reference, we reproduce here the three-point coupling of the neutral Higgs bosons with two charged Higgs bosons
\begin{equation*}
\includegraphics[viewport = 0 0 180 130,height=1.3cm,valign=c]{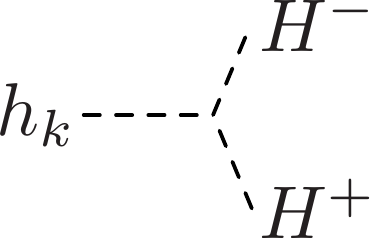} =
 -iv\lambda_{kH^+H^-}
\end{equation*}
where
\begin{equation}\label{eq:kHHcoupling}
\lambda_{kH^+H^-} = q_{k1} Z_3 + \text{Re}(Z_7 q_{k2})\,,
\end{equation}
which appears in the final result for the EDM.

In the mass-eigenstate basis, the Yukawa Lagrangian governing the coupling of Higgs fields $h_k$ and $H^\pm$ to the SM fermions $f$ is
\begin{multline}\label{eq:electronYuk}
\mathcal{L}_\text{Yuk} = \\
-\frac{m_f}{v}\sum_k h_k\bar{f}\big[q_{k1}-2T_3^f c_f \text{Re}(q_{k2})
 + i\,c_f \text{Im}(q_{k2})\gamma_5\big]f\\
 -\sqrt{2}\Big[H^+ \bar{f}' \big(\frac{m_{f'} c_{\mathrlap{f'}}}{v}\,\,P_L + \frac{m_f c_f}{v} P_R\big) V_{f'f}\,  f + \text{c.c.}\Big],
\end{multline}
where $T_3^f=\pm \frac{1}{2}$ is the third component of weak isospin, and $V_{f'f}$ is a CKM matrix element for quarks and the Kronecker delta for leptons.  The coupling coefficients $c_f$ are controlled by the $\mathbb{Z}_2$ charges assigned to the quarks and leptons.  The possible assignments yield the four 2HDM types:
\begin{align}
\label{eq:coup-type1}\text{Type~I} &:\enspace c_d = c_\ell = \cot\beta\,,\\[2mm]
\label{eq:coup-type2}\text{Type~II} &:\enspace c_d = c_\ell = -\tan\beta\,,\\[2mm]\label{eq:leptonSpecific}
\text{Lepton Specific} &:\enspace  \left\{\begin{aligned}
c_d = {}& \cot\beta \\ c_\ell ={}& -\tan\beta\,,\end{aligned}\right.\\[2mm]\label{eq:Flipped}
\text{Flipped} &:\enspace  \left\{\begin{aligned}
c_d = {}& -\tan\beta \\ c_\ell ={}& \cot\beta \,,\end{aligned}\right.
\end{align}
and $c_u = -\cot\beta$ for all types.


\section{Background field evaluation}\label{Sec:BGF-eval}
The electron EDM, $d_e$, is derived from the $q^2=0$ limit of the CP-odd Pauli form factor in the electromagnetic vertex function
\begin{equation}
\includegraphics[height=37pt,valign=c]{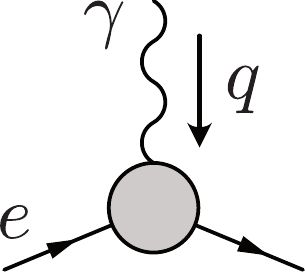} \enspace\supset\enspace i d_e \bar{u}(\mathbf{p}')\sigma^{\mu\nu}q_\nu \gamma_5 u(\mathbf{p})\,.
\end{equation}
The unsuppressed contributions to the electron EDM in the C2HDM start at two loop order.  In what follows, we present the leading order behavior of the EDM in the asymptotic limit $m_e \rightarrow 0$,  adopt a normalization that sets the overall scale
\begin{align}\label{eq:edm-norm}
\frac{d_e}{e} &= \frac{\sqrt{2} \alpha G_F m_e}{64\pi^3} \delta_e \\
\nonumber &\approx (6.5 \times 10^{-28} \,\text{cm}) \times \delta_e \,,
\end{align}
where we used $\alpha(m_Z) \approx 1/129$, and report our results in terms of the dimensionless electric dipole moment, $\delta_e$\,.

\begin{table}[t]
\centering
\begin{tabular*}{\columnwidth}{c@{\extracolsep{\fill}}ccc}
\hline
& Fermion & Charged & Gauge boson\\
Barr-Zee& loop & Higgs loop & loop \\
\hline
Electromagnetic & \multirow{2}{*}{$\delta_f^\text{EM}$ (\ref{eq:emBZ-fermion})} & \multirow{2}{*}{$\delta_{H^+}^\text{EM}$ (\ref{eq:emBZ-chargedHiggs})} & \multirow{2}{*}{$\delta_W^\text{EM}(\xi)$ (\ref{eq:emBZ-W})} \\
\includegraphics[height=20pt]{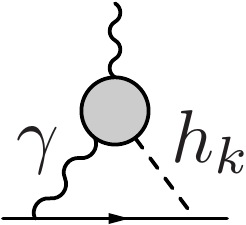}\\
Neutral current & \multirow{2}{*}{$\delta_f^\text{NC}$ (\ref{eq:ncBZ-fermion})}  & \multirow{2}{*}{$\delta_{H^+}^\text{NC}$ (\ref{eq:ncBZ-chargedHiggs})} & \multirow{2}{*}{$\delta_W^\text{NC}(\xi)$ (\ref{eq:ncBZ-W})} \\
\includegraphics[height=20pt]{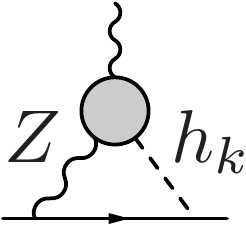}\\
Charged current & \multirow{2}{*}{--} & \multirow{2}{*}{$\delta_{H^+}^\text{CC}$ (\ref{eq:ccBZ-chargedHiggs})} & \multirow{2}{*}{$\delta_W^\text{CC}(\xi)$ (\ref{eq:ccBZ-W})} \\
\includegraphics[height=20pt]{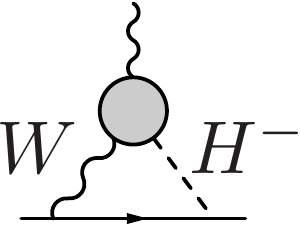} \\
\hline
Kite\\
\hline
Neutral current & \multirow{2}{*}{--} & \multirow{2}{*}{--} & \multirow{2}{*}{$\delta_\text{kite}^\text{NC}$ (\ref{eq:ncKite})}\\
\includegraphics[height=20pt]{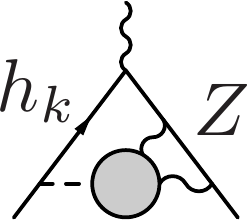} \\
Charged current & \multirow{2}{*}{--} & \multirow{2}{*}{--} & \multirow{2}{*}{$\delta_\text{kite}^\text{CC}(\xi)$ (\ref{eq:ccKite})}\\
\includegraphics[height=20pt]{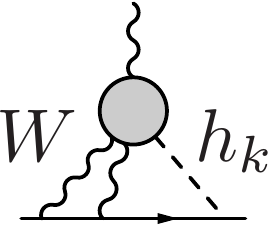} \\
\hline
\end{tabular*}
\caption{Two loop contributions to the electron EDM at $\mathcal{O}(\alpha G_F m_e)$ in the C2HDM in the background field gauge, organized by \textbf{rows:} couplings to the main lepton line and \textbf{columns:} virtual particle in the loop. Numbers in parenthesis indicate the equation number where the corresponding expression may be found.}
\label{tab:diagrams}
\end{table}

Before presenting the results of our calculation, we briefly review relevant aspects of the background field method. In the background field method, the electromagnetic vector potential is shifted in the Lagrangian to its background field value $\bar{A}_\mu(x)$ corresponding to the classical electric field coupled to the electron EDM.  Terms linear in the quantum field $A_\mu$ incurred by this shift are cancelled by a suitable choice for the  source $J^\mu_\text{em}(x)$.  In passing to the quantum theory, we choose the background field gauge condition \cite{Denner:1994xt}
\begin{multline}\label{eq:BFGaugeDef}
\mathcal{L} =  -\frac{1}{2\xi}\Big[(\partial^\mu A_\mu)^2 + \big(\partial^\mu  Z_\mu + \xi m_Z G^0\big)^2 \\
 + 2\big|(\partial^\mu + i e \bar{A}^\mu)W_\mu^+ - i \xi m_W G^+ \big|^2\Big]\,,
\end{multline}
which generalizes the conventional \tHooft $R_\xi$ gauges (see (\ref{eq:tHooftGaugeDef}) below) by maintaining covariance with respect to gauge transformations of the background field $\bar{A}_\mu$.  Compared to the conventional \tHooft $R_\xi$ gauges, the background field gauge modifies the tree-level triple gauge vertex
\begin{equation}\label{eq:bgf-tripgauge}
\Delta\left(\includegraphics[viewport = -10 0 220 150,height=1.4cm,valign=c]{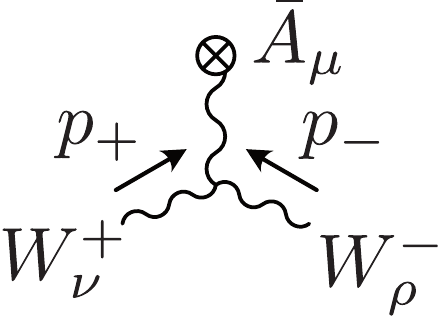}\right)
 =  \frac{-ie}{\xi}\big(g^{\mu\nu}  p_-^\rho + g^{\rho\mu} p_+ ^\nu\big)\,,
\end{equation}
includes a gauge-ghost four-point vertex
\begin{equation}\label{eq:bgf-gaugeghost}
\includegraphics[viewport = 0 0 140 130,height=1.3cm,valign=c]{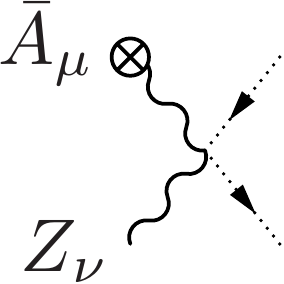}
 = \frac{ie^2\cw}{\sw} g^{\mu\nu}\,,
\end{equation}
and features the absence of the $\bar{A}_\mu$-induced $W$ gauge-Goldstone transition, substantially reducing the number of contributing Feynman diagrams.  For this reason, we provide a detailed account of our results in the background field gauge, and only provide an outline of the calculation in the conventional \tHooft $R_\xi$ gauge in section \ref{Sec:FeynmanGauge}.

With the help of \textsc{FeynArts} \cite{Hahn:2000kx}, we generated all possible two loop diagrams for the electromagnetic vertex function.  Table \ref{tab:diagrams} organizes the diagrams that contribute to the electron EDM in the background field gauge.  Groups of non-vanishing diagrams that trivially sum to zero are not shown, but are briefly mentioned in \mbox{Sec. \ref{Sec:FeynmanGauge}} in the context of the Feynman-\tHooft gauge in which they do contribute.  The Barr-Zee diagrams in the first three rows form the largest class, and are defined by containing insertions of one-loop three-point vertex functions inside the electron form factor.  Traditionally, these contributions have been classified according to the kind of three-point function that enters into the Barr-Zee diagram (rows of Table \ref{tab:diagrams}).  However considerations of gauge-invariance and scaling in the decoupling limit suggest that it is more natural to group them by degrees of freedom entering in the loop, (columns of Table \ref{tab:diagrams}).  The remaining diagrams (which we call ``kite diagrams'') are shown in the last two rows of Table \ref{tab:diagrams}, and make up a smaller set of diagrams.  Nevertheless, they formally contribute at the same order, and their inclusion is essential for gauge-independence of the final result.

In our calculations, we dimensionally regulated all Feynman integrals, and employed a naively anticommuting definition of $\gamma_5$ in the Dirac algebra.  As the EDM is UV finite to the order we work, no ambiguities associated with this definition arise.  We made extensive use of an in-house version of \textsc{Package-X} \cite{Patel:2015tea} to automate the evaluation of the two loop Feynman integrals.  In the results below, we express the contributions in terms of squared mass ratios with respect to the $k$-th neutral Higgs:  $r_k = m_f^2/m_k^2$, $w_k = m_W^2/m_k^2$, $z_k = m_Z^2/m_k^2$, and $h_k = \mHp^2/m_k^2$.  We also make frequent use of the Davydychev-Tausk vacuum integral function \cite{Davydychev:1992mt}
\begin{multline}\label{eq:davydychevtauskPhi1}
\Phi(x,y)=\text{Re}\bigg\{\frac{2}{\sqrt{\lambda}}\Big[\frac{\pi^2}{6}-\frac{1}{2}\ln x \ln y\\
+\ln\Big(\frac{1+x-y-\sqrt{\lambda}}{2}\Big)\ln\Big(\frac{1-x+y-\sqrt{\lambda}}{2}\Big)\\
-\Li\Big(\frac{1+x-y-\sqrt{\lambda}}{2}\Big)-\Li\Big(\frac{1-x+y-\sqrt{\lambda}}{2}\Big)\Big]\bigg\}\,,
\end{multline}
where $\lambda = (1-x-y)^2 - 4 x y$ is the K\"{a}ll\'{e}n polynomial, and $\text{Li}_2$ is the dilogarithm function.  The special equal-mass case is given by
\begin{align}\label{eq:davydychevtauskPhi2}
\nonumber \Phi(x) ={}& \Phi(x,x) \\
\nonumber ={}& \frac{2}{\sqrt{1-4x}}\Big[\frac{\pi^2}{6}+\ln^2\Big(\frac{1-\sqrt{1-4x}}{2}\Big)\\
{}& -\frac{\ln^2 x}{2}-2\,\Li\Big( \frac{1-\sqrt{1-4x}}{2}\Big)\Big].
\end{align}

\paragraph{Fermion loop contributions.}
\begin{figure}[t]
\includegraphics[viewport = 0 0 260 235,height=1.9cm]{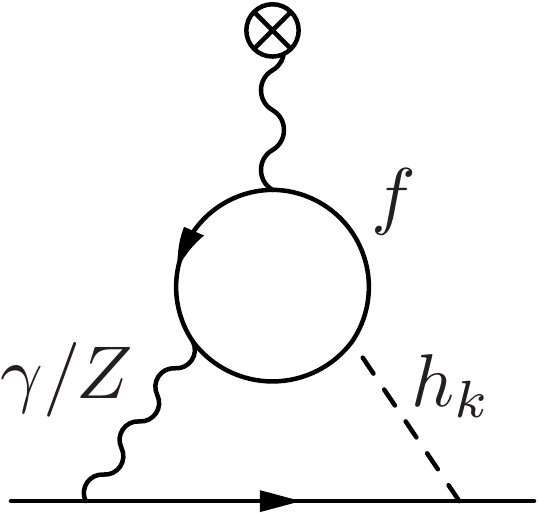}
\caption{
Representative fermion loop contribution to electromagnetic $\delta_f^\text{EM}$ (photon exchange) and neutral current $\delta_f^\text{NC}$ ($Z$ exchange) Barr-Zee diagrams.  The symbol `$\otimes$' denotes the background electromagnetic field $\bar{A}_\mu$.  Additional diagrams are obtained by reflections along the vertical axis, or by exchanging the $\gamma/Z$ and $h_k$ lines attached to the external electron.}
\label{fig:fermionBarrZee}
\end{figure}
The contributions with a fermion $f$ in the loop are shown in Fig.~\ref{fig:fermionBarrZee}, and give gauge-independent results.  The four electromagnetic Barr-Zee diagrams were originally considered in \cite{Barr:1990vd} and are given by
\begin{widetext}
\begin{equation}\label{eq:emBZ-fermion}
\begin{aligned}
\delta_f^\text{EM} ={}& - 4 N_C^f (\Qem^f)^2 \Qem^\ell \sum_k \text{Im}(q_{k2}) \Big\{c_f \big(q_{k1}-2T_3^\ell c_\ell \text{Re}(q_{k2})\big) r_k \Phi(r_k) \\
&+ \big(q_{k1}-2T_3^fc_f \text{Re}(q_{k2})\big) c_\ell  r_k \Big[4+2\ln(r_k)+(1-2 r_k \big) \Phi(r_k)\Big] \Big\}\,,
\end{aligned}
\end{equation}
where $N_C^f = 3$ for quarks and $N_C^f = 1$ for leptons, and $\Qem^f$ and $T_3^f = \pm \frac{1}{2}$ are the electric charge and third component of weak isospin, respectively.  The four neutral current diagrams give\begin{equation}\label{eq:ncBZ-fermion}
\begin{aligned}
\delta_f^\text{NC} ={}& -\frac{N_C^f \Qem^f \Qw^f \Qw^\ell}{4 \cw^2 \sw^2}\sum_k \text{Im}(q_{k2}) \Big\{c_f \big(q_{k1}-2T_3^\ell c_\ell\text{Re}(q_{k2})\big) \frac{r_k}{1-z_k} \Big(\Phi(r_k) - \sPhi{r_k}{z_k}\Big) \\
&  +\big(q_{k1}-2T_3^f c_f \text{Re}(q_{k2})\big)  c_\ell  \frac{r_k}{1-z_k}
\Big(2  \ln(z_k) + (1-2 r_k)\Phi(r_k) - \big(1 - \frac{2r_k}{z_k}\big)\sPhi{r_k}{z_k}\Big)\Big\}\,,
\end{aligned}
\end{equation}
where $\sw = \sin(\theta_W)$, $\cw = \cos(\theta_W)$, and $\Qw^f = 2T_3^f - 4 \Qem^f \sw^2$ is the weak charge of fermion $f$.

All fermion species should be added to obtain the complete contribution to the EDM.  Practically, it suffices to only include the third generation fermions $t$, $b$ and $\tau$, since other fermion contributions are suppressed by their much smaller masses.  For the lighter fermions, $b$ and $\tau$, it may be more convenient to expand $\delta_f^\text{EM}$ and $\delta_f^\text{NC}$ in small fermion masses, which can be obtained with the help of the small-argument expansion of the Davydychev-Tausk function (\ref{eq:davydychevtauskPhi2})
\begin{equation}
\Phi(x) = \Big(\ln^2(x) + \frac{\pi^2}{3}\Big) + 2 x\Big(\ln^2(x) + 2\ln(x) +\frac{\pi^2}{3}-2\Big) + \mathcal{O}(x^2)\,.
\end{equation}
\newpage
\paragraph{Charged Higgs loop contributions.}
\begin{figure*}[t]
\includegraphics[viewport = 0 0 260 235,height=1.9cm]{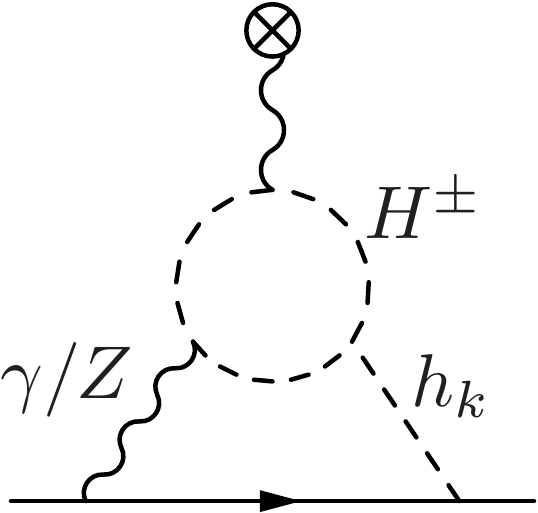}\enspace\enspace
\includegraphics[viewport = 0 0 260 235,height=1.9cm]{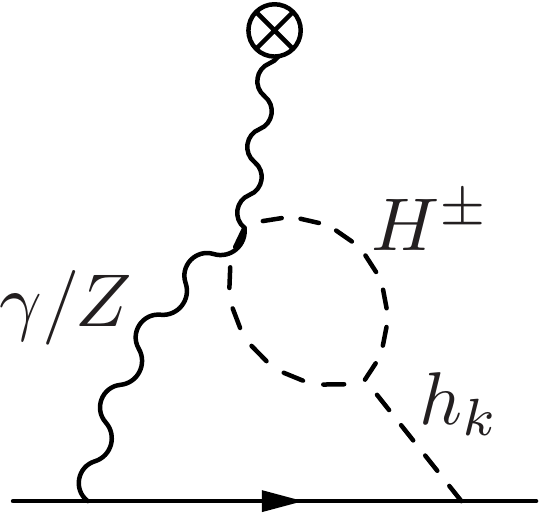}\hspace{2cm}
\includegraphics[viewport = 0 0 260 235,height=1.9cm]{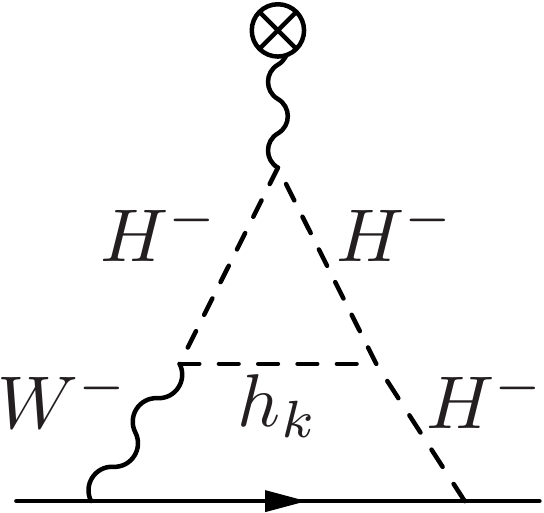}\enspace\enspace
\includegraphics[viewport = 0 0 260 235,height=1.9cm]{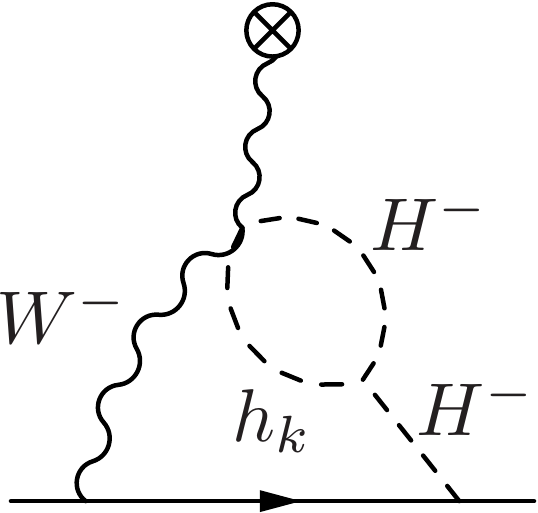}\enspace\enspace
\includegraphics[viewport = 0 0 260 235,height=1.9cm]{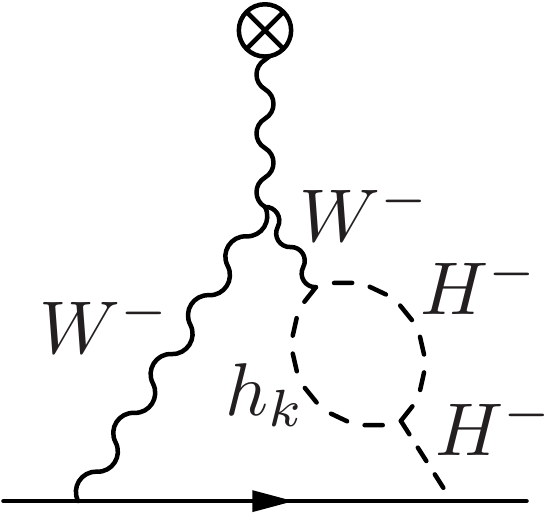}
\caption{Representative charged Higgs contributions to \textbf{left:} electromagnetic $\delta_{H^+}^\text{EM}$ and neutral current $\delta_{H^+}^\text{NC}$ Barr-Zee diagrams, and \textbf{right:} charged current $\delta_{H^+}^\text{CC}$ Barr-Zee diagrams.}
\label{fig:chargedHiggsBarrZee}
\end{figure*}
Representative Feynman diagrams involving charged Higgs loops are shown in Fig.~\ref{fig:chargedHiggsBarrZee}.  Like the fermion loop contributions, these are all gauge-independent.  For the electromagnetic Barr-Zee diagrams we find
\begin{equation}\label{eq:emBZ-chargedHiggs}
\delta_{H^+}^\text{EM} = \frac{2 \Qem^\ell \sw^2}{\pi\alpha} c_\ell \sum_k \text{Im}(q_{k2}) \lambda_{kH^+H^-}\, w_k \Big[2 + \ln(h_k) - h_k \Phi(h_k)\Big]\,,
\end{equation}
where $\lambda_{kH^+H^-}$ is the triple Higgs coupling given (\ref{eq:kHHcoupling}).
The neutral current Barr-Zee diagrams give a result proportional to $\Qw^\ell$:
\begin{equation}\label{eq:ncBZ-chargedHiggs}
\delta_{H^+}^\text{NC} =  \frac{\Qw^\ell c_{2\textsc{w}}}{4\pi\alpha} c_\ell \sum_k \text{Im}(q_{k2}) \lambda_{kH^+H^-}\frac{z_k}{1-z_k}\Big[\ln(z_k) - h_k \Phi(h_k) + \frac{h_k}{z_k}\sPhi{h_k}{z_k}\Big]\,,
\end{equation}
where $c_{2\textsc{w}} = \cos(2\theta_W)$.  Finally, for the charged current Barr-Zee diagrams we find
\begin{equation}\label{eq:ccBZ-chargedHiggs}
\begin{aligned}
\delta_{H^+}^\text{CC} = {}&  \frac{(-2T_3^\ell)}{4\pi\alpha} c_\ell \sum_k \text{Im}(q_{k2}) \lambda_{kH^+H^-}\,\Big[2 - \frac{2}{h_k} + \frac{2 \ln (h_k)}{h_k} - \frac{2-2h_k + w_k}{h_k - w_k}\ln\Big(\frac{h_k}{w_k}\Big)\\
& - \frac{1+h_k^2 - h_k(2 + w_k)}{w_k(h_k-w_k)}\ln(h_k)\ln\Big(\frac{h_k}{w_k}\Big)
- \frac{2(h_k-2h_k^2+h_k^3+w_k-2h_kw_k)}{h_k^2 w_k}\LiOM{1}{h_k} \\
&+ \frac{w_k(1- 4h_k + 2h_k^2)}{h_k^2 (h_k - w_k)}\Phi(h_k)
- \frac{1-h_k^3 - w_k + h_k^2 (3+2 w_k) - h_k(3+w_k + w_k^2)}{w_k(h_k-w_k)}\Phi(h_k,w_k)\Big].
\end{aligned}
\end{equation}
The overall sign $-2T_3^\ell$ arises from isospin ladder operators that assemble to form the commutator $\lbrack\,T_-, T_+ \rbrack$ upon combining each charged current diagram of Fig.~\ref{fig:chargedHiggsBarrZee} with its mirror image.

\paragraph{W boson loop contributions.}
The groups of Barr-Zee diagrams with $W$ boson loop shown in Fig.~\ref{fig:WBarrZee} are the largest set contributing to the electron EDM.  The 36 electromagnetic and neutral current Barr-Zee diagrams yield the gauge-dependent expressions
\begin{align}\label{eq:emBZ-W}
\delta_W^\text{EM}(\xi) ={}& \delta_W^\text{EM} + \Qem^\ell c_\ell \sum_k \text{Im}(q_{k2}) q_{k1}\Big[2\ln(\xi) + F_\xi(w_k)\Big]\,, \\ \label{eq:ncBZ-W}
\delta_W^\text{NC}(\xi) ={}& \delta_W^\text{NC} + \frac{\Qw^\ell}{4 \sw^2} c_\ell \sum_k \text{Im}(q_{k2}) q_{k1}\Big[
 \xi (1-2\sw^2) \Phi(\xi \cw^2) + F_\xi(w_k)\Big]\,,
\end{align}
with
\begin{align}\label{eq:emBZ-W_2}
\delta_W^\text{EM} ={}& \Qem^\ell c_\ell \sum_k \text{Im}(q_{k2}) q_{k1}\Big[4(1+6 w_k)
 + 2(1+6w_k)\ln(w_k) - (3-16 w_k + 12 w_k^2)\Phi(w_k) \Big]\,, \\
\delta_W^\text{NC} ={}& \frac{\Qw^\ell}{4 \sw^2} c_\ell \sum_k \text{Im}(q_{k2}) q_{k1}\Big[-\frac{3-16 w_k + 12 w_k^2}{1-z_k}\Phi(w_k)
+ \frac{1-2\sw^2+2(5-6\sw^2)w_k}{\cw^2(1-z_k)}\ln(z_k) \nonumber \\ \label{eq:ncBZ-W_2}
& \hspace{96pt}- \frac{(1+8\sw^2-12 \sw^4)z_k}{1-z_k}\Phi(\cw^2) \Big]\,.
\end{align}
Gauge-dependence is contained within the mass-dependent function,
\begin{equation}\label{eq:gaugeF}
\begin{aligned}
F_\xi(w_k) ={}&- \xi\ln^2(\xi) - 2(1-\xi)^2w_k\text{Li}_2(1-\xi) - [3+\xi-(1-\xi)^2w_k]\ln(\xi)\ln(w_k) \\
&\qquad + \xi (1-2\xi w_k)\Phi(\xi w_k) + \big[3-\xi - 2(2-\xi-\xi^2)w_k + (1-\xi)^3 w_k^2\big]\Phi(w_k,\xi w_k)\,.
\end{aligned}
\end{equation}
The result for the charged current Barr-Zee diagrams with the $W$ boson in loop is more complicated because of the presence of another mass scale from the charged Higgs.  The 12 diagrams give
\begin{equation}\label{eq:ccBZ-W}
\delta_W^\text{CC}(\xi) = \delta_W^\text{CC} + \frac{(-2T_3^\ell)}{4 \sw^2} c_\ell \sum_k \text{Im}(q_{k2}) q_{k1} G_\xi(w_k)\,,
\end{equation}
\begin{equation}\label{eq:ccBZ-W_2}
\begin{aligned}
\delta_W^\text{CC} ={}& \frac{(-2T_3^\ell)}{4 \sw^2} c_\ell \sum_k \text{Im}(q_{k2}) q_{k1}\Big[
\frac{2}{w_k}-\frac{2(1-w_k)^2}{h_k w_k}-\frac{2(1-w_k)w_k^2+(2+w_k)h_k^2-h_k(2-w_k-7w_k^2)}{h_k w_k (h_k-w_k)}\ln(w_k)\\
&+\frac{h_k^2-2(1-w_k)^2+h_k(1+7w_k)}{h_k(h_k-w_k)}\ln(h_k)
- \frac{(1-w_k)^3-3h_k^2 w_k - h_k(1+3w_k-4w_k^2)}{h_k^2(h_k-w_k)}\ln\Big(\frac{h_k}{w_k}\Big)\ln(w_k)\\
&- \frac{2w_k(1-w_k)^3+h_k(2-8w_k+6w_k^3)}{h_k^2 w_k^2}\LiOM{1}{w_k}
-\frac{1-6w_k+6 w_k^2+4 w_k^3}{(h_k-w_k)w_k^2}\Phi(w_k) \\
&+ \frac{(1-w_k)^4 - 3h_k^3 w_k - h_k(2+5w_k)(1-w_k)^2+h_k^2(1+7w_k^2)}{h_k^2(h_k-w_k)} \Phi(h_k, w_k)
\Big]\,,
\end{aligned}
\end{equation}
where
\begin{equation}\label{eq:gaugeG}
\begin{aligned}
G_\xi(w_k) ={}&
 -2\xi\big(1+\ln(\xi)\big) +\Big(\frac{(1-\xi w_k)^2}{w_k^2}-\xi\Big)\Big[\ln(\xi)\ln(\xi w_k)+2\,\LiOM{1}{\xi w_k}\Big] \\
&  - \Big[\xi(1-3\xi)-\frac{1-(1+3\xi)w_k}{w_k^2}+(1-\xi)^2\xi w_k\Big]\Phi(w_k, \xi w_k)
\end{aligned}
\end{equation}
is another mass-dependent $\xi$-dependent function.
\begin{figure*}[t]
\includegraphics[viewport = 0 0 270 235,height=1.9cm]{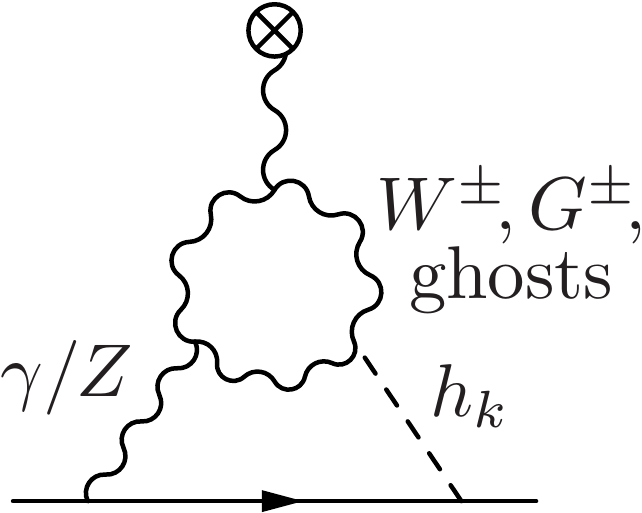}\enspace\enspace\enspace\enspace
\includegraphics[viewport = 0 0 260 235,height=1.9cm]{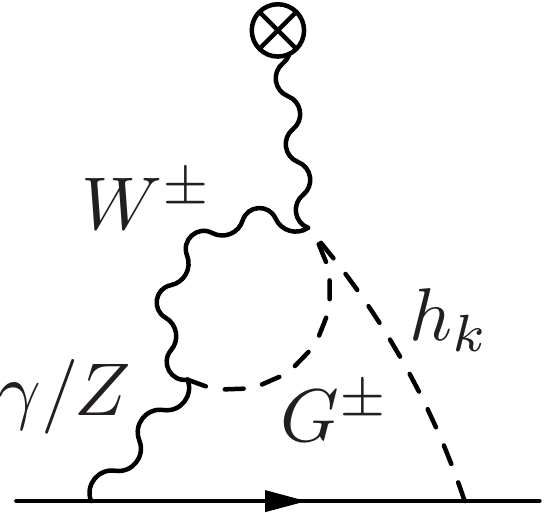}\enspace\enspace
\includegraphics[viewport = 0 0 260 235,height=1.9cm]{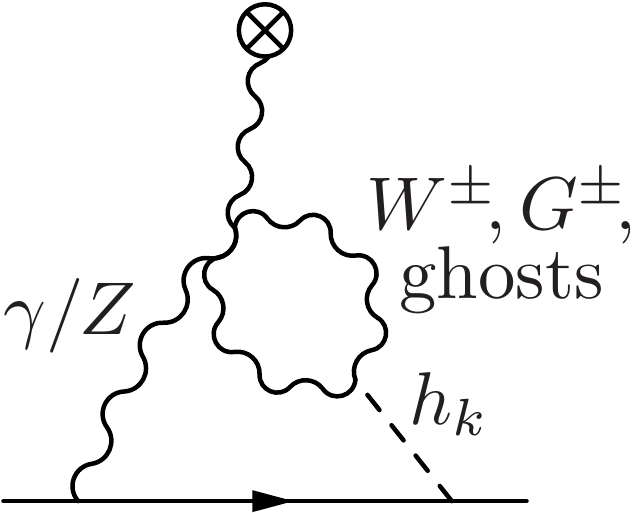}\hspace{1.5cm}
\includegraphics[viewport = 0 0 270 235,height=1.9cm]{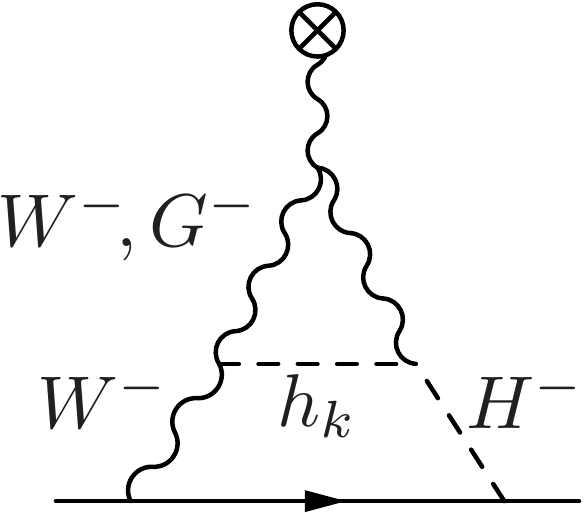}\enspace\enspace
\includegraphics[viewport = 0 0 260 235,height=1.9cm]{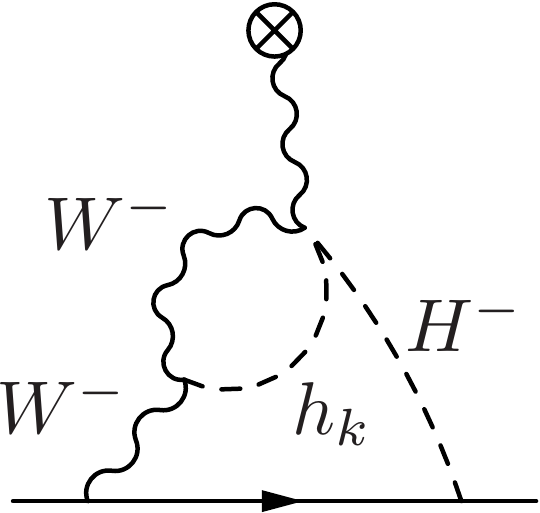}\enspace\enspace
\includegraphics[viewport = 0 0 260 235,height=1.9cm]{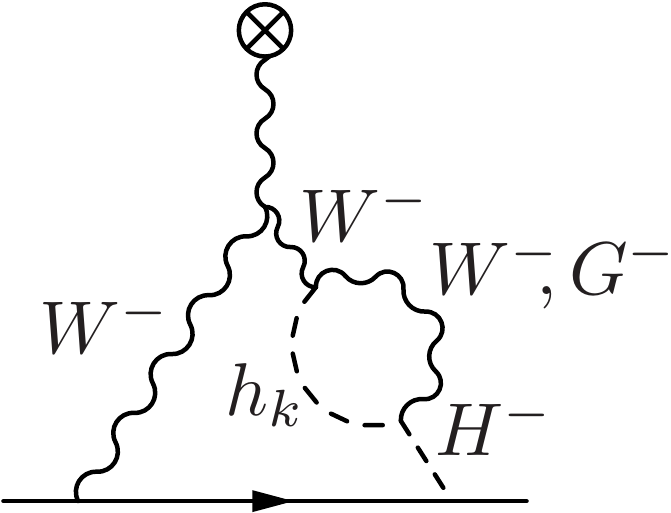}
\caption{
Representative $W$ boson contributions to \textbf{left:} electromagnetic $\delta_W^\text{EM}(\xi)$ and neutral current $\delta_W^\text{NC}(\xi)$ Barr-Zee diagrams, and \textbf{right:} charged current $\delta_W^\text{CC}(\xi)$ Barr-Zee diagrams.  Diagrams involving the 3-point coupling of the background field $\bar{A}_\mu$ to one $W$ gauge boson and one charged Goldstone boson are absent in the background field gauge.  The third diagram with a ghost loop involving the four-point coupling in Eq.~(\ref{eq:bgf-gaugeghost}) is specific to the background field gauge.}
\label{fig:WBarrZee}
\end{figure*}
\paragraph{Kite contributions.}
\begin{figure*}[b]
\includegraphics[viewport = 0 0 260 235,height=1.9cm]{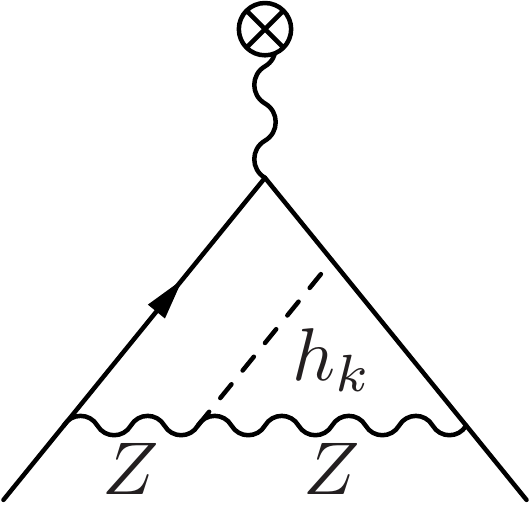}\enspace\enspace
\includegraphics[viewport = 0 0 260 235,height=1.9cm]{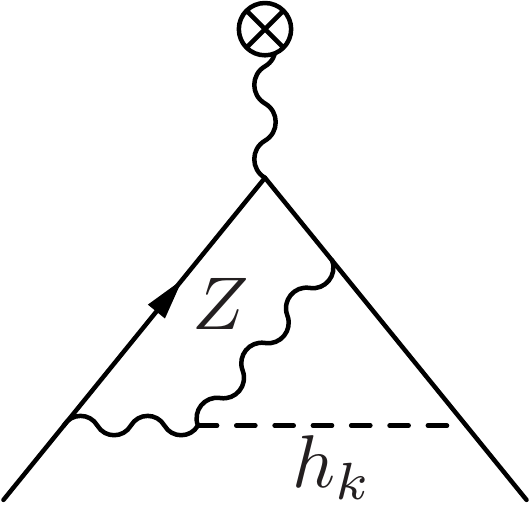}\enspace\enspace
\includegraphics[viewport = 0 0 260 235,height=1.9cm]{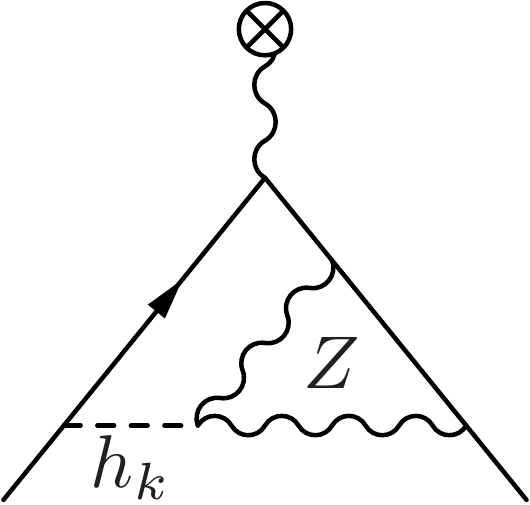}\hspace{2cm}
\includegraphics[viewport = 0 0 260 235,height=1.9cm]{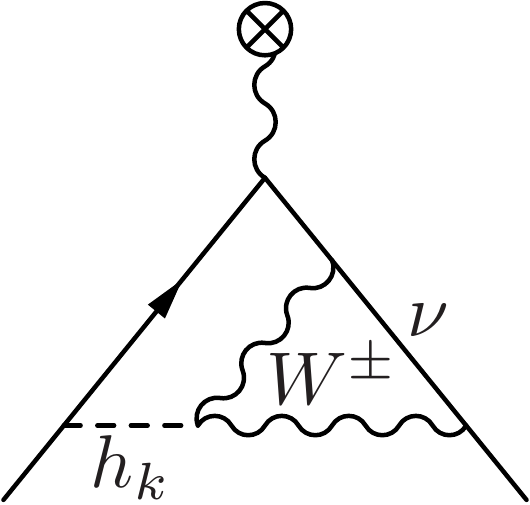}\enspace\enspace
\includegraphics[viewport = 0 0 260 235,height=1.9cm]{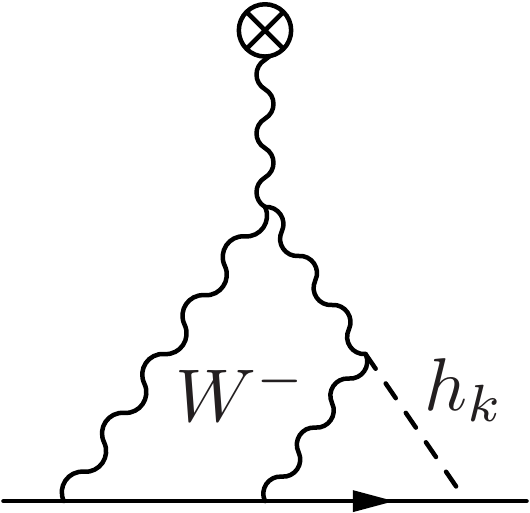}\enspace\enspace
\includegraphics[viewport = 0 0 260 235,height=1.9cm]{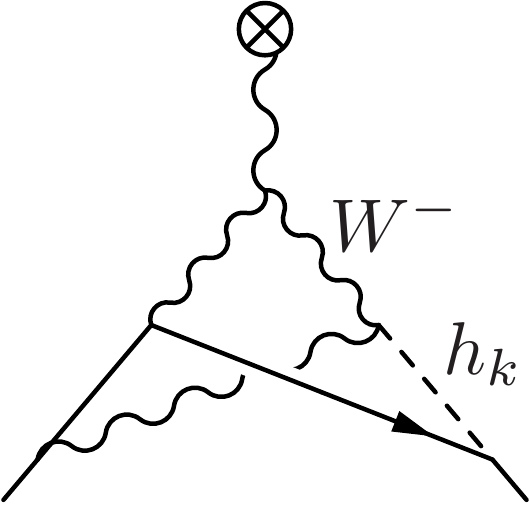}
\caption{
Representative contributions to \textbf{left:} neutral current kite $\delta_\text{kite}^\text{NC}$ and \textbf{right:} charged current $\delta_\text{kite}^\text{CC}(\xi)$ kite diagrams.}
\label{fig:kiteDiagrams}
\end{figure*}
Representative kite Feynman diagrams are shown in Fig.~\ref{fig:kiteDiagrams}.
The neutral current contribution does not depend on the gauge parameter $\xi$ and, in contrast to the neutral current Barr-Zee contributions, it is not suppressed by the weak charge $\Qw^\ell$.  In agreement with \cite{Altmannshofer:2015qra}, we find 
\begin{equation}\label{eq:ncKite}
\begin{aligned}
\delta_\text{kite}^\text{NC} ={}& -\Qem^\ell \frac{(\Qw^\ell)^2-1}{8 \sw^2 \cw^2} c_\ell \sum_k \text{Im}(q_{k2}) q_{k1}\frac{1}{z_k^3}\Big[z_k^2 + \frac{\pi^2}{6}(1-4 z_k) - 2 z_k^2 \ln(z_k) + \frac{1-4z_k}{2}\ln^2(z_k)\\
&\qquad+2(1-4z_k+z_k^2)\LiOM{1}{z_k}+\frac{1-6z_k+8z_k^2}{2}\Phi(z_k)
\Big] \\
&-\Qem^\ell \frac{(\Qw^\ell)^2+1}{24 \sw^2 \cw^2} c_\ell \sum_k \text{Im}(q_{k2}) q_{k1}\frac{1}{z_k}\Big[2z_k(1-4z_k) +\frac{\pi^2}{3}(3z_k^2+4z_k^3)-2 z_k (1+4z_k)\ln(z_k)\\
&\qquad+2(1-3z_k^2-4z_k^3) \LiOM{1}{z_k} +(1-2z_k-8z_k^2)\Phi(z_k)
\Big]\,,
\end{aligned}
\end{equation}
The charged current kite contribution is gauge-dependent, and is given by
\begin{equation}\label{eq:ccKite}
\delta_\text{kite}^\text{CC}(\xi) = \delta_\text{kite}^\text{CC} + \frac{(-2T_3^\ell)}{4 \sw^2} c_\ell \sum_k \text{Im}(q_{k2}) q_{k1}\Big[F_\xi(w_k)-G_\xi(w_k) + (1+\Qem^\ell)H_\xi(w_k)\Big]\,,
\end{equation}
with
\begin{equation}\label{eq:ccKite_2}
\begin{aligned}
\delta_\text{kite}^\text{CC} = {}& \frac{(-2T_3^\ell)}{4 \sw^2} c_\ell \sum_k \text{Im}(q_{k2}) q_{k1}\Big[\frac{2\pi^2}{9}w_k(3+4 w_k)+\frac{2}{3}(5-8w_k)-\frac{16}{3}(1+w_k)\ln(w_k) \\
&+ \frac{2(3+2w_k-6w_k^3-8w_k^4)}{3w_k^2}\LiOM{1}{w_k}+ \frac{(1+2w_k)(3-10w_k+w_k^2)}{3w_k^2}\Phi(w_k) \Big]\,.
\end{aligned}
\end{equation}
In addition to depending on $F_\xi(w_k)$ and $G_\xi(w_k)$ that appear in the $W$-loop Barr-Zee diagrams, it also involves a third $\xi$-dependent function $H_\xi(w_k)$ whose functional form is not needed since it drops out upon setting $\Qem^\ell = -1$.  This completes the listing of contributions to the electron EDM.
\end{widetext}

\paragraph{Assembling a gauge-independent result.}
Adding together the contributions listed above, the electron EDM is given by
\begin{multline}\label{eq:totaledm}
\frac{d_e}{e} = \frac{\sqrt{2}\alpha G_F m_e}{64 \pi^3}\times\\
\big[\sum_f(\delta^\text{EM}_f+\delta^\text{NC}_f) + (\delta_{H^+}^\text{EM}
+ \delta_{H^+}^\text{NC} + \delta_{H^+}^\text{CC})\\
+ (\delta_{W}^\text{EM}(\xi) + \delta_{W}^\text{NC}(\xi)  +
\delta_{W}^\text{CC}(\xi)  + \delta_\text{kite}^\text{NC} + \delta_\text{kite}^\text{CC}(\xi) )\big]\,,
\end{multline}
where we have grouped the various contributions based on the columns of Table \ref{tab:diagrams}, corresponding to the virtual particles in the loop. Gauge-dependence is contained within the Barr-Zee $W$-loop contributions and charged current kite contributions.   See Fig.~\ref{fig:gaugedependence} below for a plot of these contributions as a function of the gauge parameter.  The sum of these gauge-dependent terms yields
\begin{multline}\label{eq:gaugedep}
\delta_W^\text{EM}(\xi) + \delta_W^\text{NC}(\xi) + \delta_W^\text{CC}(\xi) + \delta_\text{kite}^\text{CC}(\xi) \Big|_\text{$\xi$-dep.} = \\
\frac{1}{4 \sw^2} c_\ell \sum_k \text{Im}(q_{k2}) q_{k1}\Big[ (\Qw^\ell -2T_3^\ell + 4 \Qem^\ell \sw^2) F_\xi(w_k)\\
-2 T_3^\ell (1+\Qem^\ell) H_\xi(w_k)+ 8 \Qem^\ell \sw^2 \ln(\xi) \\
+\Qw^\ell(1-2\sw^2)\Phi(\xi \cw^2) \Big]\,,
\end{multline}
where the $\xi$-dependent function $G_\xi(w_k)$ immediately cancels between the charged current Barr-Zee $\delta_W^\text{CC}(\xi)$ and kite $\delta_\text{kite}^\text{CC}(\xi)$ contributions.  Upon inserting the electroweak relation $\Qw^\ell = 2T_3^\ell - 4 \Qem^\ell \sw^2$ and $\Qem^\ell = -1$, the first and second terms in square brackets proportional to mass-dependent functions $F_\xi(w_k)$ and $H_\xi(w_k)$ vanish.  The remaining mass-independent terms vanish after summing over $k$, and using the orthogonality relation $\sum_k q_{1k}q_{2k} = 0$ in (\ref{eq:qorth}).  Therefore, all gauge dependent terms in (\ref{eq:totaledm}) may be safely dropped so that our final result for the electron EDM is
\begin{multline}\label{eq:gaugeinv-totaledm}
\frac{d_e}{e} = \frac{\sqrt{2}\alpha G_F m_e}{64 \pi^3}\times\\
\big[\sum_f(\delta^\text{EM}_f+\delta^\text{NC}_f) + (\delta_{H^+}^\text{EM}
+ \delta_{H^+}^\text{NC} + \delta_{H^+}^\text{CC})\\
+ (\delta_{W}^\text{EM} + \delta_{W}^\text{NC}  +
\delta_{W}^\text{CC}  + \delta_\text{kite}^\text{NC} + \delta_\text{kite}^\text{CC} )\big]\,,
\end{multline}
with the individual contributions given in (\ref{eq:emBZ-fermion}), (\ref{eq:ncBZ-fermion}), (\ref{eq:emBZ-chargedHiggs}), (\ref{eq:ncBZ-chargedHiggs}), (\ref{eq:ccBZ-chargedHiggs}), (\ref{eq:emBZ-W_2}), (\ref{eq:ncBZ-W_2}), (\ref{eq:ccBZ-W_2}), (\ref{eq:ncKite}), and (\ref{eq:ccKite}).  Despite their appearance, we emphasize that one should not interpret each component of the $k$-sum in these expressions as literally the individual contributions of the neutral Higgs to the EDM since each one by itself is gauge-dependent.  Only the sum is gauge-independent.

\section{Re-evaluation in the Feynman-\tHooft gauge}\label{Sec:FeynmanGauge}
Despite simplifications afforded by working in the background field gauge, it is still common practice to perform calculations of this kind in the conventional \tHooft $R_\xi$ gauge defined by
\begin{multline}\label{eq:tHooftGaugeDef}
\mathcal{L} =  -\frac{1}{2\xi}\Big[(\partial^\mu A_\mu)^2 + \big(\partial^\mu  Z_\mu + \xi m_Z G^0\big)^2 \\
 + 2\big|\partial^\mu W_\mu^+ - i \xi m_W G^+ \big|^2\Big]\,,
\end{multline}
and with $\xi=1$ for simplicity.  In order to facilitate comparison with earlier calculations of the EDM \cite{Leigh:1990kf,Abe:2013qla}, and also to provide additional validation of our result, we re-evaluated the electron EDM in the \tHooft $R_\xi$ gauge with $\xi$ left arbitrary.  In this section we outline how the calculation proceeds, and the steps
required to reach agreement with the background field evaluation presented above.

The electromagnetic and neutral current Barr-Zee contributions with a fermion loop $\delta_{f}^\text{EM}$, $\delta_{f}^\text{NC}$, or a charged Higgs loop $\delta_{H^+}^\text{EM}$, $\delta_{H^+}^\text{NC}$, along with the neutral current kite $\delta_\text{kite}^\text{NC}$ contributions are unchanged relative to the background field gauge.  The differences are in the electromagnetic and neutral current Barr-Zee contributions with a $W$ loop, $\delta_{W}^\text{EM}(\xi)$, $\delta_{W}^\text{NC}(\xi)$, and in the charged current contributions, $\delta_{H^+}^\text{CC}$, $\delta_{W}^\text{CC}(\xi)$ and $\delta_\text{kite}^\text{CC}(\xi)$.

Intermediate expressions are substantially more complicated due to the presence of the $\gamma W^\pm G^\mp$ vertex, which generates diagrams involving several new interaction vertices from the scalar potential.  Additionally, treatment of tadpole diagrams require a multitude of \emph{sum rules} to show that they combine with other contributions to yield a UV finite result in the end.  To avoid a barrage of lengthy expressions, we give only the parts of interest for the specific case of the Feynman-\tHooft gauge $\xi=1$.

We start with $W$ loop contributions to the electromagnetic Barr-Zee diagrams $\delta_W^\text{EM}$.  Accounting for the presence of the $\gamma W^\pm G^\mp$ vertex, there are 52 diagrams of the kinds shown on the left of Fig.~\ref{fig:WBarrZee}.  Their total is UV finite, and exhibits an apparent logarithmic singularity in the limit of vanishing electron mass
\begin{multline}
\delta_{W}^\text{EM}(\text{F'tH}) = \Qem^\ell c_\ell \sum_k \text{Im}(q_{k2}) q_{k1}\\
\times\Big[\ln\big(\frac{m_W^2}{m_e^2}\big)+\big(\text{\begin{tabular}{cc}regular as\\[-1mm]$m_e\rightarrow0$\end{tabular}}\big)\Big]\,.
\end{multline}
After performing the $k$-sum and using the orthogonality relations in (\ref{eq:qorth}), the singularity vanishes.

\begin{figure}
\includegraphics[viewport = 0 0 260 235,height=1.9cm]{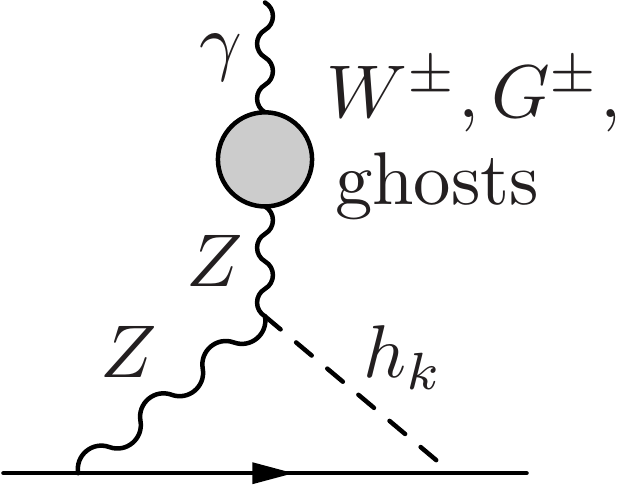}
\caption{
Class of diagrams additionally contributing to $W$ loop neutral current Barr-Zee, $\delta^{NC}_W$, in the \tHooft $R_\xi$ gauge.}
\label{fig:FeynmanGaugeNCDiagram}
\end{figure}
Next we consider the $W$ loop neutral current Barr-Zee contributions $\delta_W^\text{NC}$.  There are 52 diagrams that sum to a UV divergent expression with the pole part in $d=4-2\epsilon$ dimensions given by
\begin{equation}\label{eq:ncBZ-W-fg}
\frac{\Qw^\ell}{4 \sw^2} c_\ell \sum_k \text{Im}(q_{k2}) q_{k1}\frac{4z_k}{(1-z_k)^2}\big(1-z_k+\ln(z_k)\big)\frac{1}{\epsilon}
\end{equation}
that cannot be removed by performing the $k$-sum on account of the nontrivial $m_k^2$ dependence.
However, there is another class of diagrams to consider, shown in Fig.~\ref{fig:FeynmanGaugeNCDiagram}, involving the $\gamma$-$Z$ transition function mediated by gauge loops.  We mention that, in the background field gauge, individual diagrams in this group are non-vanishing but sum to zero because of the property that $\Pi_{\bar{A}Z}^{\mu\nu}(q^2) \rightarrow 0$ as $q^2 \rightarrow 0$ in this gauge \cite{Denner:1994xt}.  In the \tHooft $R_\xi$ gauges this group does not vanish, and importantly, it supplies a UV divergent contribution equal and opposite to (\ref{eq:ncBZ-W-fg}), yielding an overall finite neutral current contribution $\delta_{W}^\text{NC}(\text{F'tH})$.

We now report on the charged current kite contribution.  There are a total of 10 diagrams of the type shown in the last three diagrams of Fig.~\ref{fig:kiteDiagrams}, and four additional ones involving the $\gamma W^\pm G^\mp$ vertex.  Their total is nominally UV divergent
\begin{equation}
\delta_\text{kite}^\text{CC}(\text{F'tH}) = \frac{1}{4 \sw^2} c_\ell \sum_k \text{Im}(q_{k2}) q_{k1}\Big[-\frac{1}{2\epsilon} + \text{finite}\Big]\,.
\end{equation}
But after performing the $k$-sum, the UV divergent part vanishes by orthogonality of the rotation vectors (\ref{eq:qorth}).

Despite their finiteness, none of the three contributions $\delta_{W}^\text{EM}(\text{F'tH})$, $\delta_{W}^\text{NC}(\text{F'tH})$, nor $\delta_\text{kite}^\text{CC}(\text{F'tH})$ so far considered coincide with their background field gauge counterparts.  To find agreement, the charged current contributions $\delta_{H^+}^\text{CC}$ and $\delta_{W}^\text{CC}$ need to be examined, which we now do.

\begin{figure}[t]
\begin{tabular*}{0.57\columnwidth}{@{\extracolsep{\fill} } cc }
\includegraphics[viewport = 0 0 260 235,height=1.9cm]{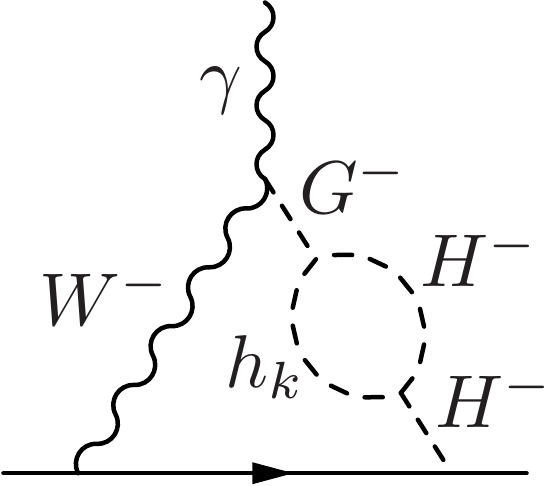}&
\includegraphics[viewport = 0 0 260 235,height=1.9cm]{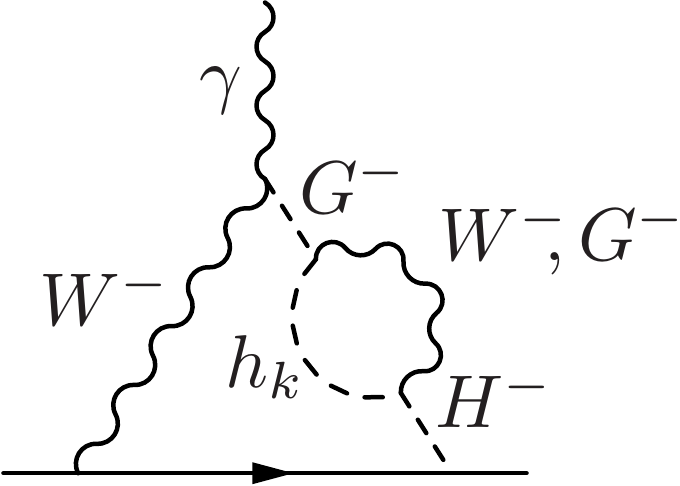}\\
\textbf{a}&\textbf{b}
\end{tabular*}
\caption{
Additional charged current Barr-Zee diagrams in the \tHooft $R_\xi$ gauge.  The $R$-subtracted finite parts  of diagrams (a) and (b) contribute to $\delta_{H^+}^\text{CC}$ and $\delta_{W}^\text{CC}$, respectively.  The UV-singular $R$-subtractions cancel against the tadpole diagrams in Fig.~\ref{fig:Tadpole1FeynmanGauge} and \ref{fig:Tadpole2FeynmanGauge}.}
\label{fig:FeynmanGaugeCCDiagrams}
\end{figure}

The analysis of charged current contributions and their separation into $\delta_{H^+}^\text{CC}$ and $\delta_{W}^\text{CC}$ appears at first obfuscated by numerous diagrams that must be considered in addition to those shown on the right of Figs.~\ref{fig:chargedHiggsBarrZee} and \ref{fig:WBarrZee}. A little investigation shows that to recover the charged current contributions, we only need to include the $R$-subtracted part of the diagrams in Fig.~\ref{fig:FeynmanGaugeCCDiagrams} (and their mirror images).  The $R$-subtractions contain the UV singular parts of these diagrams stemming from the sub-loop Goldstone-Higgs transition function. In $d$ dimensions, these are given by
\begin{multline}\label{eq:rSubH}
\delta_R\big[\text{Fig.~\ref{fig:FeynmanGaugeCCDiagrams}(a)}\big] = \frac{c_\ell}{d\,\sw^2}\mathbf{\Delta}(m_W,m_{H^+}) \, \sum_k \text{Im}(q_{k2})\\
\times
v^2 \lambda_{k H^+ H^-} \big(\mathbf{A}_0(\mHp) - \mathbf{A}_0(m_k)\big)
\end{multline}
and
\begin{multline}\label{eq:rSubW}
\delta_R\big[\text{Fig.~\ref{fig:FeynmanGaugeCCDiagrams}(b)}\big] = \frac{c_\ell}{d\,\sw^2}\mathbf{\Delta}(m_W,m_{H^+}) \, \sum_k \text{Im}(q_{k2}) \\
\times q_{k1}\big(m_k^2\mathbf{A}_0(m_W) + (\mHp^2 \! -m_k^2)\mathbf{A}_0(m_k)\big)\,,
\end{multline}
where $\mathbf{A}_0(m)$ and $\mathbf{\Delta}(m_W,m_{H^+})$ are the one-loop tadpole and triangle integrals defined by
\begin{gather}
\begin{aligned}
\mathbf{A}_0(m) &= \int(dk) \frac{1}{k^2-m^2},\\
\mathbf{\Delta}(m_W,m_{H^+}) &= \int(dk) \frac{1}{(k^2-m_W^2)^3(k^2-\mHp^2)}\,.
\end{aligned}
\end{gather}

Then upon adding the six Barr-Zee diagrams of the type shown to the right of Fig.~\ref{fig:chargedHiggsBarrZee} to the $R$-subtracted form of Fig.~\ref{fig:FeynmanGaugeCCDiagrams}(a), we obtain a UV finite charged current charged Higgs loop contribution that also agrees with the corresponding background field gauge evaluation given in Eq.~(\ref{eq:ccBZ-chargedHiggs}),
\begin{equation}
\delta_{H^+}^\text{CC}(\text{F'tH}) = \delta_{H^+}^\text{CC}.
\end{equation}

Similarly, by adding the 16 Barr-Zee diagrams of the type shown on the right of Fig.~\ref{fig:WBarrZee} to the $R$-subtracted forms of Fig.~\ref{fig:FeynmanGaugeCCDiagrams}(b), we obtain a UV finite result for the charged current $W$ loop contribution $\delta_W^\text{CC}(\text{F'tH})$.  Finally, upon combining this to the electromagnetic, neutral current Barr-Zee diagrams and the charged current kite contributions in the \tHooft $R_\xi$ gauge computed above, we obtain a result precisely equal to the sum of corresponding contributions in the background field gauge
\begin{multline}
\delta_W^\text{EM}(\text{F'tH}) + \delta_W^\text{NC}(\text{F'tH}) + \delta_W^\text{CC}(\text{F'tH}) + \delta_\text{kite}^\text{CC}(\text{F'tH}) = \\
\delta_W^\text{EM}(\xi) + \delta_W^\text{NC}(\xi) + \delta_W^\text{CC}(\xi) + \delta_\text{kite}^\text{CC}(\xi).
\end{multline}
To confirm the equivalence analytically, and especially to demonstrate $\xi$-independence, we found it essential to expand the \tHooft $R_\xi$ gauge results into partial fractions with respect to $m_k^2$ and to perform the $k$-sum dispensing of any parts that vanish by orthogonality of the rotation vectors $q_{k1}$ and $q_{k2}$.

\begin{figure}[t]
\begin{tabular*}{0.89\columnwidth}{@{\extracolsep{\fill} } ccc }
\includegraphics[viewport = 0 0 260 235,height=1.9cm]{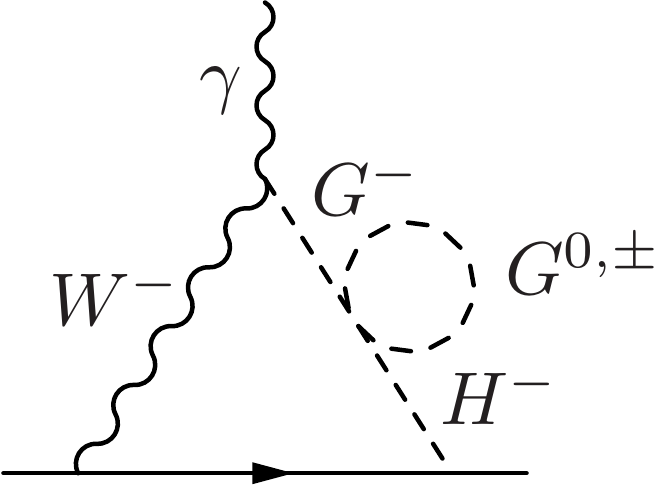}&
\includegraphics[viewport = 0 0 260 235,height=1.9cm]{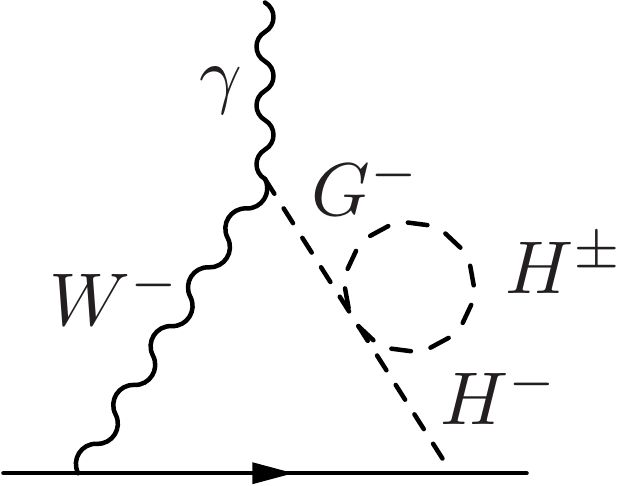}&
\includegraphics[viewport = 0 0 260 235,height=1.9cm]{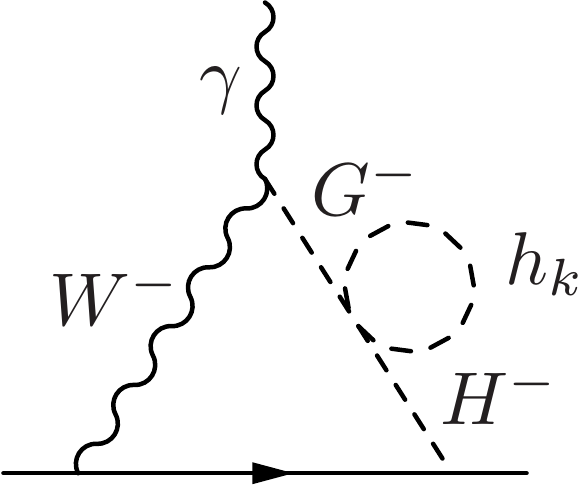}\\
\textbf{a}&\textbf{b}&\textbf{c}
\end{tabular*}
\caption{Diagrams involving the Goldstone-Higgs transition function that contribute to the electron EDM in the \tHooft $R_\xi$ gauge.}
\label{fig:Tadpole1FeynmanGauge}
\end{figure}

\begin{figure}[t]
\begin{tabular*}{0.89\columnwidth}{@{\extracolsep{\fill} } ccc }
\includegraphics[viewport = 0 0 260 235,height=1.9cm]{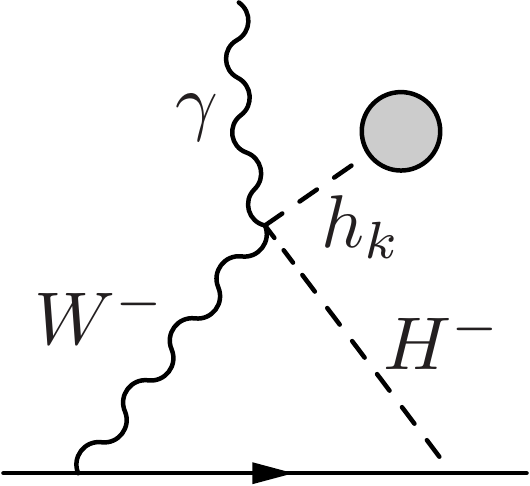}&
\includegraphics[viewport = 0 0 260 235,height=1.9cm]{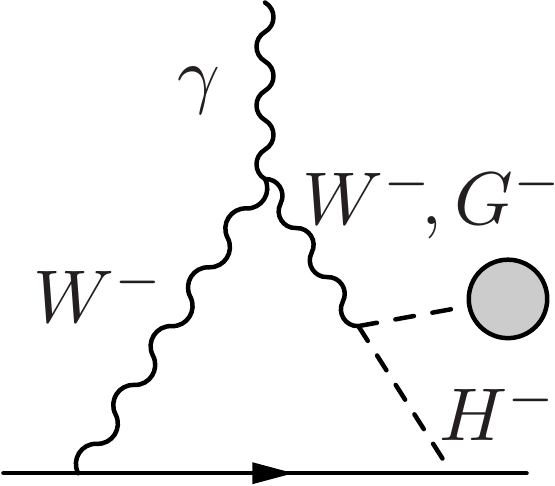}&
\includegraphics[viewport = 0 0 260 235,height=1.9cm]{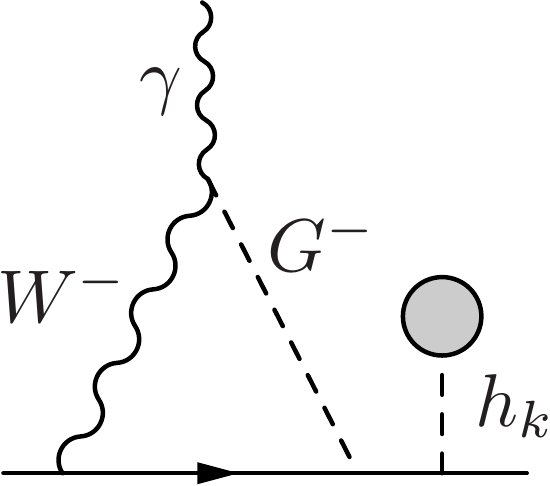}\\
\textbf{a}&\textbf{b}&\textbf{c}
\end{tabular*}
\caption{Tadpole diagrams in the \tHooft $R_\xi$ gauge.  Diagram (c) represents a contribution due to a CP-violating shift in the residue of the electron pole.}
\label{fig:Tadpole2FeynmanGauge}
\end{figure}

Finally, we turn to the remaining diagrams shown in Fig.~\ref{fig:Tadpole1FeynmanGauge} and \ref{fig:Tadpole2FeynmanGauge}.  We aim to demonstrate a cancellation between these diagrams and the $R$-subtractions of Fig.~\ref{fig:FeynmanGaugeCCDiagrams}, given by (\ref{eq:rSubH}) and (\ref{eq:rSubW}).
Diagrams (a) and (b) of Fig.~\ref{fig:Tadpole1FeynmanGauge} are unusual in that the neutral Higgs bosons are absent and hence do not involve a $k$-sum.  Furthermore, they depend on four-point interaction vertices from the scalar potential
\begin{equation*}
\includegraphics[viewport = -10 0 170 150,height=1.45cm,valign=c]{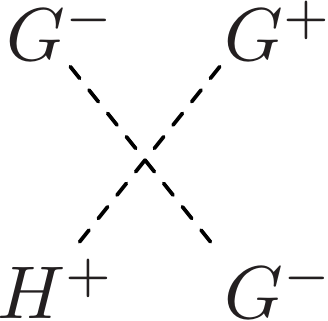} = \,\,2\left(\includegraphics[viewport = -10 0 160 150,height=1.45cm,valign=c]{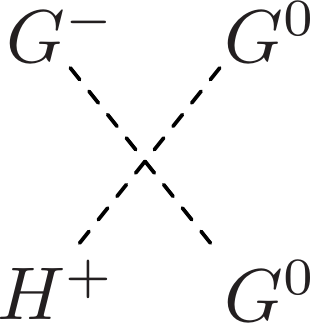}\right) = -2 i Z_6\,,
\end{equation*}
and
\begin{equation*}
\includegraphics[viewport = -10 0 180 150,height=1.45cm,valign=c]{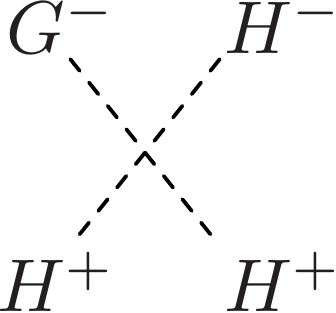} = -2 i Z_7\,,
\end{equation*}
that so far have not appeared in this calculation.  To put these contributions under a $k$-sum so that they may be brought together with other diagrams, we replace $Z_6$ and $Z_7$ by their sum rules
\begin{align}
\label{eq:sr1}Z_6 &= \frac{1}{v^2}\sum_k q_{k2}^* q_{k1} m_k^2,\\
\label{eq:sr2}Z_7 &= \sum_k q_{k2}^* \lambda_{k H^+ H^-}\,.
\end{align}
Respectively, these are derived by considering the double contraction of the diagonalized neutral Higgs squared-mass matrix $(R \mathcal{M}^2 R^\top)_{jk}$ in (\ref{eq:DiagMassMtx}) with $q_{k2}^* q_{j1}$, and the contraction of the triple Higgs coupling $\lambda_{k H^+ H^-}$ in (\ref{eq:kHHcoupling}) with $q_{k2}^*$.  The diagram in Fig.~\ref{fig:Tadpole1FeynmanGauge}(c) involves the four-point coupling
\begin{gather*}
\includegraphics[viewport = -10 0 160 150,height=1.45cm,valign=c]{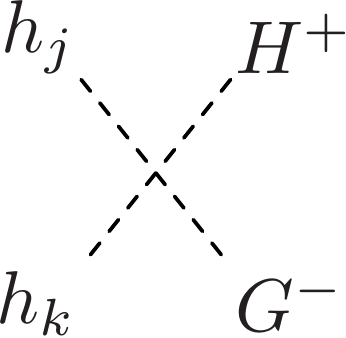} = - i \lambda_{jkH^+G^-}\,,
\end{gather*}
whose diagonal elements are given by
\begin{equation}\label{eq:fourpointscalar}
\lambda_{kkH^+G^-} = q_{k1} \big(q_{k2}^* Z_4 +  q_{k2} Z_5 + q_{k1} Z_6 \big) + |q_{k2}|^2 Z_7\,.
\end{equation}
Together, the diagrams of Fig.~\ref{fig:Tadpole1FeynmanGauge} yield
\begin{multline}\label{eq:tad1}
\big[\text{Fig.~\ref{fig:Tadpole1FeynmanGauge}}\big] = \frac{-c_\ell}{d\,\sw^2} \mathbf{\Delta}(m_W,m_{H^+}) \sum_k \Big\{\text{Im}(q_{k2}) \\
\times\Big[q_{k1}m_k^2 \big(2\mathbf{A}_0(m_W)
+ {\textstyle\frac{1}{2}} \mathbf{A}_0(m_Z) \big) \\
+ 2 v^2 \lambda_{k H^+ H^-} \mathbf{A}_0(\mHp)  \Big]\\
+\frac{v^2}{2} \text{Im}(\lambda_{kk H^+ G^-}) \mathbf{A}_0(m_k)\Big\}.
\end{multline}

Next, we consider the tadpole diagrams of Fig.~\ref{fig:Tadpole2FeynmanGauge}.  In the background field gauge, diagrams (a) and (b) cancel tadpole-by-tadpole on account of the triple-gauge vertex (\ref{eq:bgf-tripgauge}).   In the Feynman-\tHooft gauge, however, these diagrams give the non-zero result
\begin{multline}\label{eq:tad2-ab}
\big[\text{Fig.~\ref{fig:Tadpole2FeynmanGauge}(a, b)}\big] = \frac{c_\ell}{d\,\sw^2}  \sum_k \text{Im}(q_{k2}) T_k \\
\times\Big[\mathbf{\Delta}(m_W,m_{H^+}) + (4-d)(2-d)\smash{\frac{\mathbf{A}_0(m_W)}{m_W^4 m_k^2}}\Big]\,,
\end{multline}
where
\begin{multline}
T_k = - 4 \sum_f N_C^f \big(q_{k1} - 2 T_3^f c_f \text{Re}(q_{k2})\big) m_f^2 \mathbf{A}_0 (m_f) \\
+ q_{k1} \big(2 (d-1) m_W^2 + m_k^2\big) \mathbf{A}_0(m_W) \\
+ q_{k1} \big((d-1) m_Z^2 + {\textstyle\frac{1}{2}}m_k^2 \big) \mathbf{A}_0(m_Z)  \\
+ v^2 \lambda_{k H^+ H^-} \mathbf{A}_0(\mHp) + \frac{v^2}{2} \sum_j \lambda_{kjj} \mathbf{A}_0(m_j)
\end{multline}
is the tadpole function to which fermions, $W$, $Z$, \text{ghosts}, $G^\pm$, $G^0$, $H^\pm$, and $h_k$ contribute.
Diagram (c) of Fig.~\ref{fig:Tadpole2FeynmanGauge} represents an EDM contribution derived from a one-loop magnetic moment contribution induced by a CP-violating shift in the residue of the electron propagator pole.  When added to diagrams (a) and (b), this contribution exactly cancels the second term in square brackets of (\ref{eq:tad2-ab}).  Then, after performing the $k$-sum, all contributions to $T_k$ proportional to $q_{k1}$ and $q_{k2}$ but independent of $m_k^2$ drop out by orthogonality, leaving just the Goldstones, charged Higgs, and neutral Higgs bosons
\begin{multline}\label{eq:tad2}
\big[\text{Fig.~\ref{fig:Tadpole2FeynmanGauge}}\big] = \frac{c_\ell}{d\,\sw^2}\mathbf{\Delta}(m_W,m_{H^+})\sum_k \text{Im}(q_{k2}) \\
\times\Big[q_{k1}m_k^2 \big(\mathbf{A}_0(m_W)
+ {\textstyle\frac{1}{2}} \mathbf{A}_0(m_Z) \big) \\
+ v^2 \lambda_{k H^+ H^-} \mathbf{A}_0(\mHp) + \frac{v^2}{2} \sum_j \lambda_{kjj} \mathbf{A}_0(m_j) \Big]\,.
\end{multline}
The neutral Higgs tadpole contribution is a double sum involving the triple-Higgs vertex
\begin{gather*}
\includegraphics[viewport = 0 0 180 130,height=1.3cm,valign=c]{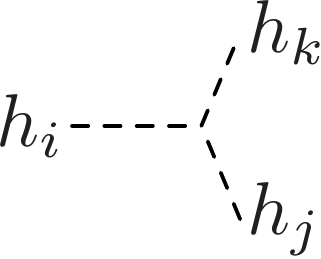} = -i v \lambda_{ijk}\,,
\end{gather*}
whose diagonal elements are given by
\begin{align}\label{eq:tripHiggs}
\nonumber \lambda_{kjj} ={} & 3 q_{k1} q_{j1}^2 Z_1 \\
\nonumber & \enspace + \big(q_{k1}|q_{j2}|^2 + 2\,\text{Re}(q_{j1}q_{k2}q_{j2}^*)\big)(Z_3+Z_4)\\
\nonumber & \enspace + \text{Re}\big[(q_{k1}q_{j2}^2 + q_{j1}q_{k2}q_{j2})Z_5\\
\nonumber & \enspace + 3 (q_{j1}^2q_{k2} + 2 q_{j1} q_{k1} q_{j2})Z_6\\
& \enspace + (q_{k2}^* q_{j2}^2+ 2 q_{k2}|q_{j2}|^2)Z_7\big]\,.
\end{align}
To combine this result with (\ref{eq:tad1}), we perform the (outer) $k$-sum on the last term of (\ref{eq:tad2}) to exchange $\lambda_{kjj}$ for $\lambda_{jjH^+G^-}$ with the help of the sum rule
\begin{multline}
\sum_{k}q_{k2}^* \lambda_{kjj} = \lambda_{jjH^+G^-} \\[-4mm]
  + 2 q_{j2}^*\lambda_{j H^+H^-}
 - 2 q_{j2}^* q_{j1} \frac{\mHp^2\!- m^2_j}{v^2}\,,
\end{multline}
which is explicitly verified by inserting the definitions (\ref{eq:kHHcoupling}), (\ref{eq:fourpointscalar}) and (\ref{eq:tripHiggs}), and applying the orthogonality relations.  Then, upon adding (\ref{eq:tad2}) to (\ref{eq:tad1}), $Z$-Goldstone contributions and terms proportional to $\lambda_{kkH^+G^-}$ cancel yielding
\begin{multline}
\big[\text{Figs.~\ref{fig:Tadpole1FeynmanGauge}}+\text{\ref{fig:Tadpole2FeynmanGauge}}\big] = \frac{-c_\ell}{d\,\sw^2}\mathbf{\Delta}(m_W,m_{H^+})\smash{\sum_k}\,\text{Im}(q_{k2})\\
\times
\Big[q_{k1}\big(m_k^2\mathbf{A}_0(m_W) + (\mHp^2 \! -m_k^2)\mathbf{A}_0(m_k)\big)\\
+v^2 \lambda_{k H^+ H^-} \big(\mathbf{A}_0(\mHp) - \mathbf{A}_0(m_k)\big)\Big]\,,
\end{multline}
which, in turn, completely cancels the $R$-subtractions given in (\ref{eq:rSubH}) and (\ref{eq:rSubW}).  This completes our evaluation of the electron EDM in the \tHooft $R_\xi$ gauge, thereby establishing agreement with our result in the background field gauge.

\section{Light quark EDMs}\label{Sec:Lightquark}
\begin{figure}[b]
\includegraphics[viewport = 0 0 260 235,height=1.9cm]{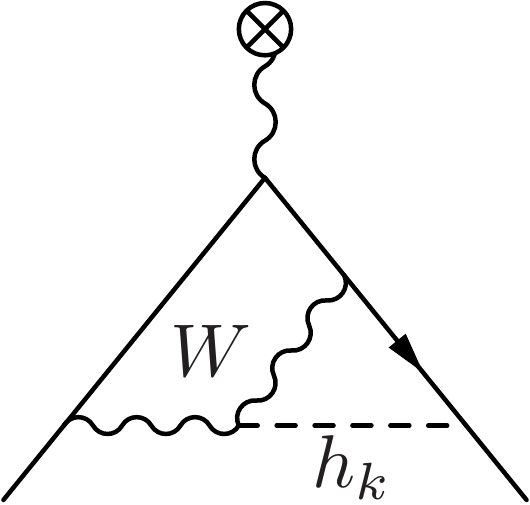}
\caption{
Charged current kite diagram that contributes to quark EDMs in the background field gauge.  Other diagrams do not contribute at $\mathcal{O}(G_F m_q)$.}
\label{fig:quarkkite}
\end{figure}
In this section, we briefly digress to discuss how our results can be adapted to obtain EDMs of light quarks.  Denoting $q$ as a generic light quark flavor, we adopt the normalization of the quark EDM $d_q$ as in (\ref{eq:edm-norm}), with the replacement $m_\ell \rightarrow m_q$.  Then, our background field gauge results (\ref{eq:emBZ-fermion})--(\ref{eq:ccKite}) should be modified by replacing the electron charges and couplings with the corresponding ones for quarks
\begin{equation}
\{\Qem^\ell, \Qw^\ell , T_3^\ell, c_\ell \} \longrightarrow \{ \Qem^q, \Qw^q , T_3^q, c_q \}\,.
\end{equation}
Also, there are new charged current kite contributions shown in Fig.~\ref{fig:quarkkite}.  Including them, and putting $\Qem^u = +2/3$ and $\Qem^d = -1/3$ in the formulae gives somewhat different results for their gauge-independent parts.  For EDMs of up and charm quarks, the expression in (\ref{eq:ccKite_2}) should be replaced by
\begin{multline}
\delta_\text{kite}^\text{CC} = \frac{(-2T_3^u)}{4 \sw^2} c_u \sum_k \text{Im}(q_{k2})q_{k1}\Big[\frac{4\pi^2}{27}w_k(3+4 w_k)\\
+\frac{2}{9}(13-16w_k)-\frac{4}{9}(11+8 w_k)\ln(w_k) \\
+ \frac{2(9+4w_k-12w_k^3-16w_k^4)}{9w_k^2}\LiOM{1}{w_k}\\
+ \frac{(1+2w_k)(9-32w_k+11w_k^2)}{9w_k^2}\Phi(w_k)
\Big]\,,
\end{multline}
and for down and strange quarks, (\ref{eq:ccKite_2}) should be replaced by
\begin{multline}
\delta_\text{kite}^\text{CC} = \frac{(-2T_3^d)}{4 \sw^2} c_d \sum_k \text{Im}(q_{k2})q_{k1}\Big[\frac{2\pi^2}{27}w_k(3+4 w_k)\\
+\frac{2}{9}(11-8w_k)-\frac{8}{9}(5+2 w_k)\ln(w_k) \\
+ \frac{2(9+2w_k-6w_k^3-8w_k^4)}{9w_k^2}\LiOM{1}{w_k}\\
+ \frac{(1+2w_k)(9-34w_k+19w_k^2)}{9w_k^2}\Phi(w_k)
\Big]\,.
\end{multline}
The total quark EDM is given by (\ref{eq:gaugeinv-totaledm}) with the replacement $m_e\rightarrow m_q$.

The generalization to top and bottom quark EDMs requires a separate treatment due to their large masses and Yukawa couplings.  In practice, this means the inclusion of new classes of diagrams involving multiple Higgs exchange that are suppressed for light quarks.  Furthermore, since it is not justified to expand the Feynman integrals in small top quark mass, the calculation is technically more challenging.  For these reasons, we have not carried out the calculation.

\section{Comparison with literature}\label{Sec:comparison}
\begin{figure*}[tb]\centering
\includegraphics[width=0.8\columnwidth]{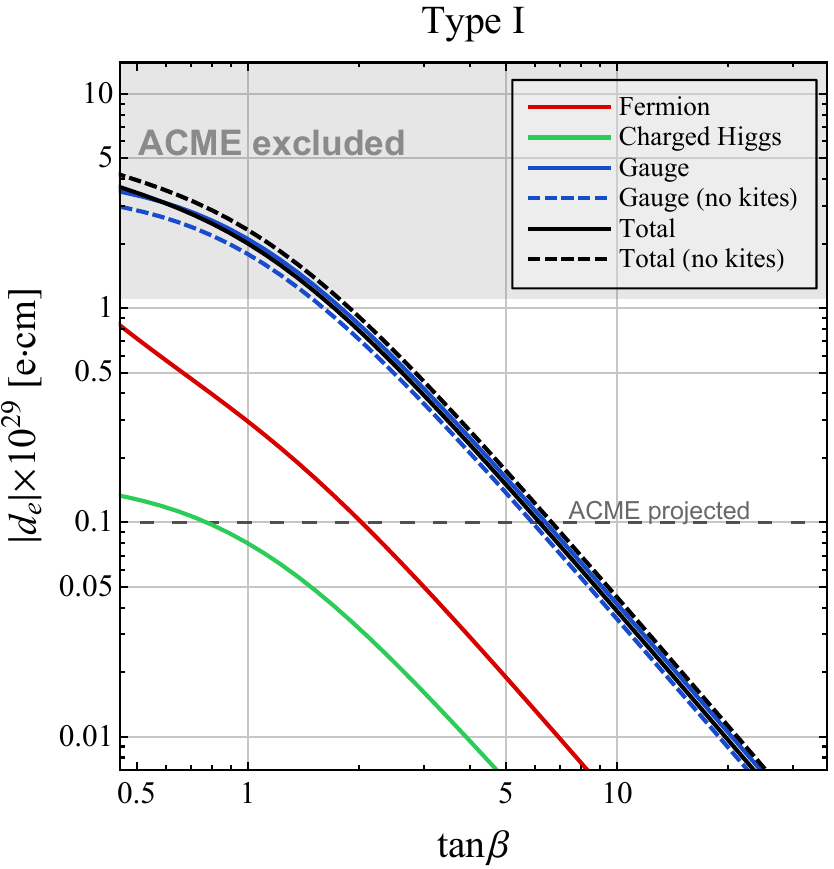}\hspace{0.1\columnwidth}
\includegraphics[width=0.8\columnwidth]{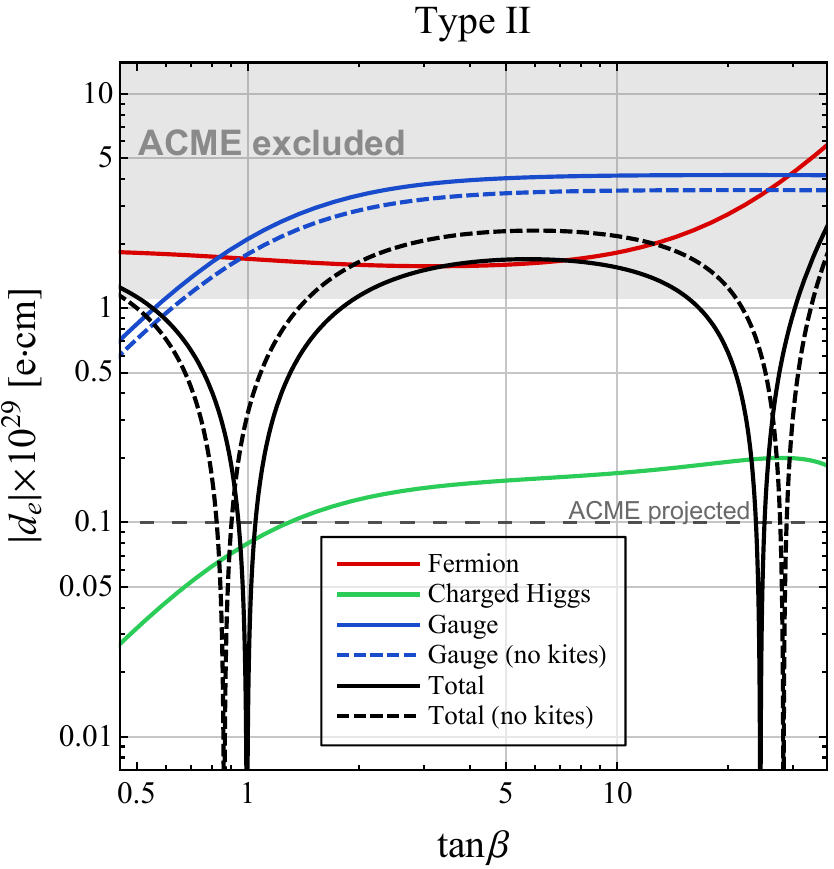}
\caption{Predictions of the electron EDM in the \textbf{left:} Type~I, and \textbf{right:} Type~II C2HDM as a function of $\tan\beta$ for the benchmark point in (\ref{eq:c2hdmParams}).  The solid black line represents the full result in (\ref{eq:gaugeinv-totaledm}).  The solid red, green, and blue curves are obtained by summing all contributions within each column of Table \ref{tab:diagrams} labeled `Fermion loop', `Charged Higgs loop', and `Gauge boson loop' respectively.  The dashed lines are the corresponding contributions without the charged and neutral current kite diagrams in the background field Feynman gauge, $\xi=1$.  The shaded region corresponds to the 90\% C.L. exclusion limit from the ACME collaboration. In the future, ACME is expected to improve the bound by at least an order of magnitude. This is indicated by the horizontal dashed line.
}
\label{fig:type1-tanbeta-plot}
\end{figure*}

The electron EDM in the C2HDM has been the subject of a long history of investigations by numerous authors, consisting of efforts to identify and calculate the important two loop contributions \cite{Barr:1990vd,Gunion:1990ce, Leigh:1990kf, Chang:1990sf}.  The original results of the gauge boson loop contributions were understood not to exhibit gauge-invariance largely due to the omission of contributions involving the charged Higgs boson or the omission of kite diagrams.  An effort was undertaken relatively recently by Abe \emph{et. al.} \cite{Abe:2013qla} to rectify the shortcomings of the earlier analyses to obtain a gauge-invariant result.  Even though this work still does not constitute a complete calculation of the electron EDM as emphasized by the authors, their results have become a standard reference for subsequent phenomenological studies involving the electron EDM in the C2HDM \cite{Cheung:2014oaa,Inoue:2014nva, Chen:2015gaa, Muhlleitner:2017dkd,Basler:2017uxn} (see also~\cite{Bian:2014zka, Bian:2016zba, Fontes:2017zfn, Keus:2017ioh,Chun:2019oix, Fuyuto:2019svr, Cheung:2020ugr, Kanemura:2020ibp, Chen:2020soj} for recent related studies). Therefore in this section, we compare our results with Abe \emph{et. al.}, and we investigate the extent to which our complete two loop result modifies predictions for the electron EDM relative to theirs.

The work of Abe \emph{et. al.} focuses on calculating all Barr-Zee contributions, with special attention to the off shell three-point functions that enter them.  They argue that in the \tHooft $R_\xi$ gauge (\ref{eq:tHooftGaugeDef}) the $W$-loop Barr-Zee contributions $\delta_W^\text{EM}$, $\delta_W^\text{NC}$, and $\delta_W^\text{CC}$ are not gauge-invariant because the three-point functions fail to exhibit transversality with respect to the off shell leg.  To obtain transverse three-point functions, they algebraically extract specific parts from the charged current kite diagrams $\delta_\text{kite}^\text{CC}$ using the electroweak pinch technique \cite{Papavassiliou:1989zd,Degrassi:1992ue,Papavassiliou:1994pr}, and add them to the Barr-Zee diagrams.  In this way, they achieve a gauge-invariant result for the electron EDM insofar as the pinch technique leads to gauge-invariant off shell Green functions.  Since results derived from the pinch technique coincide with those in the background field gauge (\ref{eq:BFGaugeDef}) with $\xi = 1$ \cite{Denner:1994nn,Hashimoto:1994ct,Pilaftsis:1996fh}, we were able to compare our results with theirs for each of the eight contributions listed in the first three rows of Table \ref{tab:diagrams}.  After careful comparison, we found exact agreement for all of them.  The remainder of the kite contributions were left unevaluated.

We now explore how our inclusion of the kite contributions numerically affects the prediction of the electron EDM.  To that end, we use the following input for the SM parameters~\cite{Zyla:2020zbs}: 
\begin{equation}\label{eq:smParams}
\begin{aligned}
m_\tau &= \phantom{00}1.777\text{ GeV} & & & m_W &= \phantom{0}80.34\text{ GeV} \\
m_b &= \phantom{00}2.88\text{ GeV} & & & m_Z &= \phantom{0}91.19\text{ GeV} \\
m_t &= 163.0\text{ GeV} & & & m_h &= 125\text{ GeV} \\
\alpha(m_Z) &= 1/129 &&& v &= 246\text{ GeV}\,,
\end{aligned}
\end{equation}
with $\cw = m_W/m_Z$.  Additionally, we fix the C2HDM parameters to the following benchmark values 
\begin{equation}\label{eq:c2hdmParams}
\begin{aligned}
\mHp &= 420\text{ GeV} & & & Z_3 &= \phantom{+}2.0 \\
\text{Im}(\lambda_5) &=\phantom{+}0.01 & & & Z_4 &= -0.45 \\
\text{Re}(Z_5) &= -1.25 & & & \text{Re}(Z_6) &= -0.001\,,
\end{aligned}
\end{equation}
and investigate the electron EDM as a function of $\tan\beta$.  Note that, as discussed in the appendix, this set of 7 parameters completely fixes the Higgs potential of the C2HDM.  The mass spectrum at this benchmark point is $\{m_1,\,m_2,\,m_3,\,m_{H^+}\}\approx \{125,\,350,\,450,\,420\}\,\text{GeV}$, and depends very mildly on $\tan\beta$.  Tree level vacuum stability is satisfied and all parameters remain perturbative at this benchmark over the interval $0.5\lesssim \tan\beta \lesssim40$.  Additionally, it leads to a phenomenology that is generally in agreement with experimental bounds \cite{workInProgress}. We mention that larger values of $\tan\beta$ for the Type~II model may already be excluded by direct searches for heavy Higgs bosons at the LHC based on the $H\to\tau\tau$ channel \cite{Sirunyan:2018zut,Aad:2020zxo}. These bounds are relaxed in the Type~I, Flipped, or Lepton Specific models. Moreover, a charged Higgs boson mass in the few hundred GeV mass range is liable to introduce sizable contributions to the $b\to s \gamma$ transition.  Ref. \cite{Misiak:2020vlo} showed that for the Type~II model, the lower limit on $\mHp$ is around 800 GeV, with mild dependence on $\tan\beta$.  But more recently, ref. \cite{Bernlochner:2020jlt} emphasized new significant theoretical uncertainties in the determination of the $b\to s\gamma$ rate, leaving more room for new physics contributions.   The corresponding bound in the Flipped 2HDM will be similar. Type~I, and Lepton Specific models will be less constrained by the $b\to s\gamma$ rate because of the $\tan\beta$ suppression of the down quark Yukawa couplings (\ref{eq:coup-type1}) and (\ref{eq:leptonSpecific}). The determination of the exact bound on $\mHp$ is beyond the scope of this paper.

Fig.~\ref{fig:type1-tanbeta-plot} shows how various contributions to the electron EDM depend on $\tan\beta$ at the benchmark point in Type I (left panel) and Type II (right panel) C2HDM.  The results for Flipped and Lepton Specific models are qualitatively similar to the ones for Type~I and Type~II models respectively, and therefore we do not show them.   Over the domain of $\tan\beta$ shown, the CP-violating component of the SM-like Higgs boson, $h_1$, is in the range $10^{-4} \lesssim|{\rm{Im}}(q_{12})|\lesssim 10^{-3}$.  The colored lines are the sums of all contributions within each column of Table \ref{tab:diagrams} as labeled in the figure.  The black line shows the total contribution to the electron EDM.  To compare with the predictions of Abe \emph{et. al.} \cite{Abe:2013qla}, we also show the result of omitting the charged and neutral current kite diagrams as dashed lines.

\begin{figure}[tb]\centering
\includegraphics[width=0.8\columnwidth]{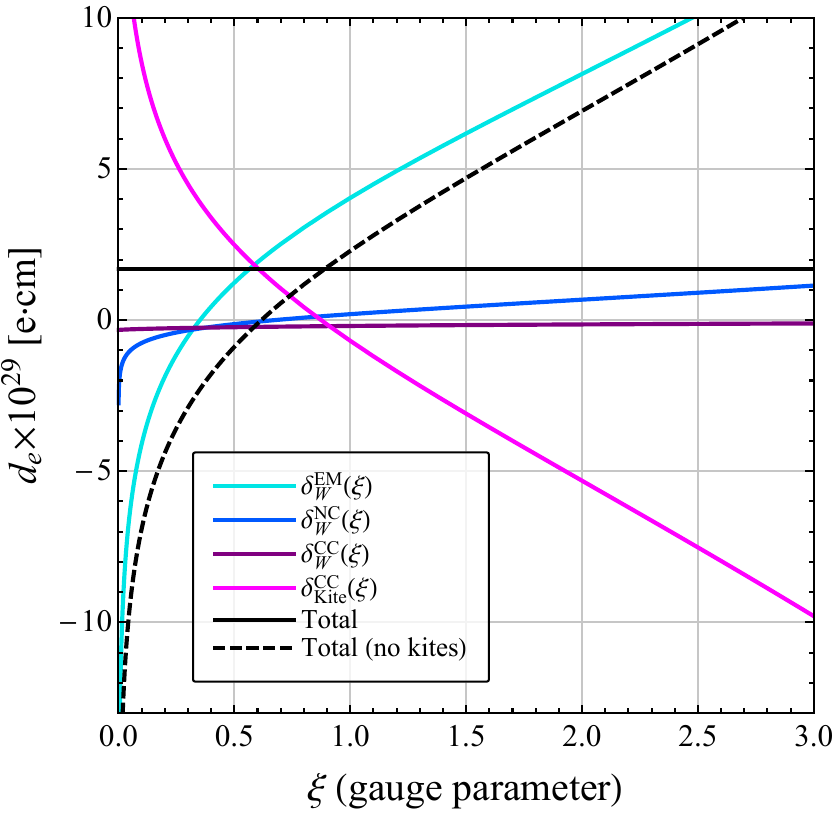}
\caption{Gauge-dependence of individual contributions to the electron EDM listed in the last column of Table~\ref{tab:diagrams} in the background field gauge for the Type~II model at the benchmark point in (\ref{eq:c2hdmParams}) with $\tan\beta=5$.
The horizontal black line is the total gauge-independent EDM in (\ref{eq:gaugeinv-totaledm}), and the dashed black curve is the total excluding the charged current $\delta_\text{kite}^\text{CC}(\xi)$ and neutral current $\delta_\text{kite}^\text{NC}$ kite contributions.}
\label{fig:gaugedependence}
\end{figure}

In the Type~I C2HDM, all contributions to the electron EDM are negative and their magnitudes fall with increasing $\tan\beta$ on account of the couplings in (\ref{eq:coup-type1}).  On the other hand, in the Type~II C2HDM, the electron coupling enters with an opposite sign and rises with $\tan\beta$ according to the couplings in (\ref{eq:coup-type2}).  This causes the charged Higgs (green curve) and gauge (blue curve) contributions to grow with increasing $\tan\beta$ and to contribute to the EDM with a positive sign.  As a result, cancellations due to destructive interference against the fermion contributions (red curve) can cause the predicted EDM to drop below the current and even future expected sensitivity of ACME in some regions.  At our benchmark point,  cancellations occur around $\tan\beta\approx 1$ and $25$.  These cancellations were first noticed and emphasized in \cite{Inoue:2014nva}.  However, the cancellations they found at larger $\tan\beta$ fall in regions of parameter space
outside the domain of perturbativity. Our findings show that cancellations are still possible in the Type~II C2HDM even when all couplings remain perturbative.

The inclusion of kite diagrams 
can lead to important numerical shifts in the prediction for the electron EDM.  This effect is particularly pronounced in the Type~II model wherein the gauge and the fermion contributions are of comparable size but enter with an opposite sign.  Including the kite diagrams leads to substantial shifts of the cancellation point in $\tan\beta$.  Furthermore, without the kite diagrams, the remaining contributions are gauge-dependent.  In Fig.~\ref{fig:gaugedependence}, we plot the individual gauge-dependent contributions $\delta_W^\text{EM}(\xi)$, $\delta_W^\text{NC}(\xi)$, $\delta_W^\text{CC}(\xi)$, and $\delta_\text{kite}^\text{CC}(\xi)$ in the background field gauge over a range of the gauge parameter $\xi$.  The horizontal black line is the gauge-independent EDM obtained by including all contributions.  The dashed black line is the EDM without the kite contributions.  It is remarkable that without the kite contributions, even a mild variation in $\xi$ can flip the sign of the EDM, highlighting the importance of a complete gauge-independent calculation.


\section{Decoupling Limit and EFT analysis}\label{Sec:Decoupling}
In this section, we consider the possibility that the new Higgs bosons of the C2HDM are very heavy ($m_{2,3}, m_{H^+} \gg v$) by investigating the asymptotic behavior of electron EDM near the decoupling limit.  We find that the electron EDM exhibits a logarithmic dependence on the heavy masses, and that its dependence on the C2HDM parameters is considerably simplified.

The decoupling limit is achieved by formally taking $Y_2\rightarrow\infty$, with all other parameters in the Higgs basis fixed \cite{Gunion:2002zf}.  To determine the asymptotic behavior of the electron EDM in this limit, we require the large $Y_2$ behavior of the mixing matrix elements $q_{k1}$, $q_{k2}$, the coupling $\lambda_{k H^+ H^-}$, and all the mass-dependent loop functions.  In this section, we rename $Y_2=M^2$ to emphasize its status as a large mass, since in this limit the additional Higgs bosons of the C2HDM collectively scale as
\begin{equation}
m_{2,3}^2 = \mHp^2 = M^2\big[1 + \mathcal{O}(\textstyle\frac{v^2}{M^2})\big]\,.
\end{equation}
The mass of the lightest Higgs boson scales as a constant
\begin{equation}
m_1^2 = Z_1 v^2\big[1 + \mathcal{O}(\textstyle\frac{v^2}{M^2})\big] \equiv m_h^2 \,,
\end{equation}
which we therefore identify as the SM Higgs mass $m_h = 125$\,GeV.  To leading order, the elements of the rotation vectors (\ref{eq:qmtx}) scale as
\begin{equation}
q_{k1} = \begin{pmatrix}1\\[1mm]\frac{v^2}{M^2}\text{Re} (Z_6 e^{-i \theta_5/2})\\[1mm]-\frac{v^2}{M^2}\text{Im} (Z_6 e^{-i \theta_5/2})\end{pmatrix},\enspace   q_{k2} = \begin{pmatrix}-\frac{v^2}{M^2}Z_6^* \\[1mm] e^{-i\theta_5/2} \\[1mm] i e^{-i\theta_5/2} \end{pmatrix},
\end{equation}
where $\theta_5 = \arg(Z_5)$, and the components of the triple Higgs coupling $\lambda_{kH^+H^-}$ in (\ref{eq:kHHcoupling}) scale as
\begin{align}
 \text{Im}(q_{k2}) \lambda_{kH^+H^-}\Big|_{k=1} &= \mathcal{O}(\textstyle\frac{v^2}{M^2}),\\
\sum_{k=2}^3 \text{Im}(q_{k2}) \lambda_{kH^+H^-} &=
- \text{Im}\big(Z_7 \big) + \mathcal{O}(\textstyle\frac{v^2}{M^2})\,.
\end{align}
To obtain the behavior of the loop functions near the decoupling limit, the $k=1$ and $k=2,3$ components of the $k$-sums over the neutral Higgs bosons need to be examined separately.  Loop functions independent of heavy masses $m_2$, $m_3$ and $\mHp$ are necessarily $\mathcal{O}(1)$, and offer no further simplification.  For loop functions containing heavy masses, we obtain the leading asymptotic behavior by directly expanding the original momentum-space Feynman integrals by regions \cite{Smirnov:1999bza}, and check the results by analytically expanding the explicit expressions manually.

Ultimately, we find that the electron EDM is proportional to $\text{Im}(Z_{6,7}) = \pm \sin\beta\cos\beta\,\text{Im}(\lambda_5)$ 
and contains a logarithmically enhanced contribution near the decoupling limit that arises from the $W$ loop Barr-Zee diagrams, yielding the leading logarithmic approximation
\begin{equation}\label{eq:leadinglogapprox}
\delta_e = \frac{-3}{4\cw^2}\frac{v^2}{M^2} c_\ell \sin\beta \cos\beta\, \text{Im}(\lambda_5)\ln\Big(\frac{M^2}{m_W^2}\Big) \, .
\end{equation}
For TeV-scale Higgs masses, this logarithm is not particularly large, and may not dominate over the non-logarithmic contributions.  In the following, we therefore provide the complete asymptotic expansion of the electron EDM through $\mathcal{O}(v^2/M^2)$.   We find it convenient to classify each contribution as either long distance, $\Delta^\text{IR}$, and short distance, $\Delta^\text{UV}$, according to an effective field theory (EFT) analysis (to be discussed shortly below) to write the EDM as
\begin{align}
\nonumber \delta_e ={}& \frac{v^2}{M^2} \sin\beta \cos\beta\, \text{Im}(\lambda_5) \times \\
\nonumber & \Big[\sum_f c_f \Delta^\text{IR}_{f\text{(P)}} + c_\ell (\sum_f \Delta^\text{IR}_{f\text{(S)}} + \Delta^\text{IR}_\text{NC kite} + \Delta^\text{IR}_W) \\
\label{eq:UVIR} &\enspace + c_\ell (\Delta^\text{UV}_W + \Delta^\text{UV}_{H^+}) + \mathcal{O}({\textstyle\frac{v^2}{M^2}}) \Big] \,.
\end{align}
In what follows, we express squared mass ratios with respect to the mass of the SM Higgs boson $r = m_f^2/m_h^2$, $w = m_W^2/m_h^2$, and $z = m_Z^2 / m_h^2$.  The contributions from fermion loop Barr-Zee diagrams give
\begin{multline}\label{eq:irFP}
\Delta^\text{IR}_{f\text{(P)}} = -4 N_C^f (\Qem^f)^2 \Qem^\ell r \,\Phi(r) \\
- \frac{N_C^f \Qem^f \Qw^f \Qw^\ell}{4 \cw^2 \sw^2} \frac{r}{1-z} \Big(\Phi(r) - \sPhi{r}{z}\Big),
\end{multline}
and
\begin{multline}\label{eq:irFS}
\Delta^\text{IR}_{f\text{(S)}} = -4 N_C^f (\Qem^f)^2 \Qem^\ell   r \Big[4 + 2\ln(r) + (1-2 r) \Phi(r)\Big]  \\
 - \frac{N_C^f \Qem^f \Qw^f \Qw^\ell}{4 \cw^2 \sw^2}  \frac{r}{1-z}
\Big(2  \ln(z)\\
 + (1-2 r)\Phi(r) -  \big(1 - \frac{2r}{z}\big)\sPhi{r}{z}\Big) \,,
\end{multline}
where `S' and `P' refer to the coupling of the Higgs boson to fermion $f$ in the loop.
The leading behavior of the neutral current kite contribution is
\begin{widetext}
\begin{equation}\label{eq:irKite}
\begin{aligned}
\Delta_\text{NC kite}^\text{IR} ={}& -\Qem^\ell \frac{(\Qw^\ell)^2-1}{8 \sw^2 \cw^2 z^3} \Big[z^2 + \frac{\pi^2}{6}(1-4 z) - 2 z^2 \ln(z) + \frac{1-4z}{2}\ln^2(z)\\
&\qquad+2(1-4z+z^2)\LiOM{1}{z}+\frac{1-6z+8z^2}{2}\Phi(z)
\Big] \\
&-\Qem^\ell \frac{(\Qw^\ell)^2+1}{24 \sw^2 \cw^2 z} \Big[2z(1-4z) +\frac{\pi^2}{3}(3z^2+4z^3)-2 z (1+4z)\ln(z)\\
&\qquad+2(1-3z^2-4z^3) \LiOM{1}{z} +(1-2z-8z^2)\Phi(z)\Big]\,.
\end{aligned}
\end{equation}
The sum of the long distance parts of the leading behavior of the $W$ loop Barr-Zee and the charged current kite diagrams is
\begin{equation}\label{eq:IRW}
\begin{aligned}
\Delta^\text{IR}_W = {}& -\frac{3}{4 \cw^2} \Big[\frac{1}{2\epsilon} -\gamma_E + \ln(4\pi) + \ln\big(\frac{\mu^2}{m_W^2}\big)+\frac{7}{4}\Big] \\
& + \frac{1}{4\sw^2} \Big\{\Big[\frac{2\pi^2}{9}w(3+4 w) + \frac{2(3+5w-(8+144\sw^2)w^2)}{3w}\\
& \qquad\qquad - \frac{2\big(3+4(2+3\sw^2)w+8(1+9\sw^2)w^2\big)}{3 w}\ln(w) + \frac{2(3+2w-6w^3-8w^4)}{3w^2}\LiOM{1}{w} \\
& \qquad\qquad + \Big(\frac{(3-16w+12 w^2)(1-4\sw^2 z)}{1-z}+\frac{3-4w-19w^2+2w^3}{3w^2}\Big)\Phi(w)\Big]\\
& \qquad + \frac{\Qw^\ell}{\cw^2}\Big[
\frac{1-2\sw^2+2(5-6\sw^2)w}{(1-z)}\ln(z) + (\cw^2-\sw^2)\ln(\cw^2) - \frac{(1+8\sw^2-12 \sw^4) w}{(1-z)}\Phi(\cw^2)\Big]\Big\}\,,
\end{aligned}
\end{equation}
\end{widetext}
whereas the short distance part is given by
\begin{equation}\label{eq:UVW}
\Delta^\text{UV}_W = \frac{3}{4\cw^2} \Big[\frac{1}{2\epsilon}  -\gamma_E + \ln(4\pi) + \ln\big(\frac{\mu^2}{M^2}\big)+\frac{7}{4}\Big]\,.
\end{equation}
Finally, the leading behavior of the charged Higgs Barr-Zee contributions is
\begin{equation}\label{eq:UVH}
\Delta^\text{UV}_{H^+} = \frac{3}{4 \cw^2} \big(\Phi(1) - 2\big)\,,
\end{equation}
where $\Phi(1) \approx 2.344$.  Observe that when (\ref{eq:IRW}) and (\ref{eq:UVW}) are added together, the parameters of dimensional regularization $1/2\epsilon + \ln(\mu^2)$ and associated constants $-\gamma_E+\ln(4\pi)+7/4$ cancel, and the leading logarithm of (\ref{eq:leadinglogapprox}) is recovered.  These unphysical parameters are introduced as a result of identifying and separating the long distance contributions derived from the Standard Model EFT, which we now discuss.

\begin{figure}
\includegraphics[height=2.0cm,valign=c]{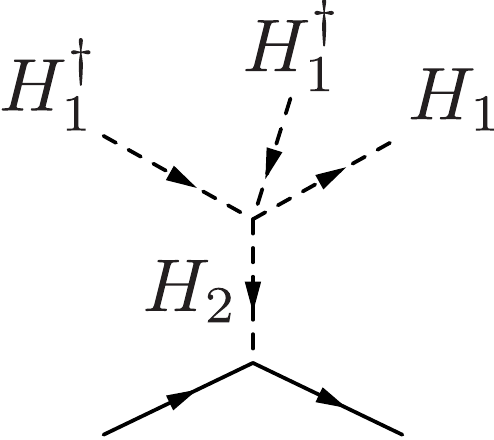} $\longrightarrow$ \enspace
\includegraphics[height=1.7cm,valign=c]{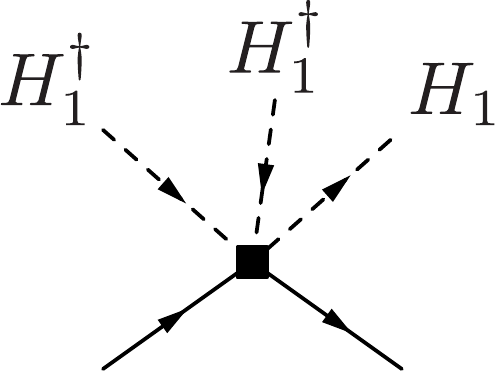}
\caption{
Generation of the CP-violating effective operator in (\ref{eq:dim6}) by integrating out $H_2$ at tree level.}
\label{fig:integratingout}
\end{figure}

The Standard Model EFT contains higher-dimensional effective operators that parametrize new physics above the electroweak scale.  In the context of the C2HDM, these operators are generated by integrating out the heavy Higgs bosons in the decoupling limit \cite{Egana-Ugrinovic:2015vgy}.  Among the CP-violating effective operators, the one relevant to the electron EDM at $\mathcal{O}(v^2/M^2)$ is the dimension-6 operator \cite{Egana-Ugrinovic:2018fpy}
\begin{equation}
\label{eq:dim6}
\mathcal{L}_6 = \frac{y_f}{M^2} c_f Z_6^* (H^\dag H) (H\bar{f}_L) f_R + \text{c.c.}\,,
\end{equation}
that arises by integrating out $H_2$ from the tree-level interaction shown in Fig.~\ref{fig:integratingout}.  Here, $y_f = \sqrt{2} m_f/v$ is the SM Yukawa coupling, $H\equiv H_1$ is the SM Higgs field, and $f_L$ and $f_R$ are the left-handed isodoublet and right-handed isosinglet fermions, respectively.  From an agnostic bottom-up point of view, the only unambiguous part of the electron EDM that can be determined from the EFT in (\ref{eq:dim6}) is the leading logarithm (\ref{eq:leadinglogapprox}).  However, since the value of the logarithm is not particularly large unless $M^2$ is far above the TeV scale, it is interesting to explore the extent to which the non-logarithmic terms of the full asymptotic behavior of the electron EDM can be reproduced in the infrared.

There are two classes of interactions derived from the operator in (\ref{eq:dim6}) in the electroweak vacuum that contribute to the electron EDM. The first class of interactions is the pseudoscalar Yukawa interaction which is obtained by setting two of the Higgs fields to their vacuum expectation values
\begin{equation}\label{eq:EFT-relevant}
  \mathcal{L}_6 \supset -i\frac{v^2}{M^2} c_f \sin\beta\cos\beta\,\text{Im}(\lambda_5) \frac{m_f}{v}  h \bar f \gamma_5 f  \,.
\end{equation}
In the background field gauge, the diagrams involving these interactions are essentially identical to those that are considered for the full C2HDM, but with those containing a charged Higgs boson omitted (Fig.~\ref{fig:fermionBarrZee}, left of Fig.~\ref{fig:WBarrZee}, and Fig.~\ref{fig:kiteDiagrams}).  We find that these contributions are UV finite as expected from power counting arguments, but also gauge-dependent.  These contributions were  calculated in \cite{Altmannshofer:2015qra} in the background field Feynman gauge, and we find agreement when we set $\xi=1$ in our formulas.

\begin{figure}[t]
\includegraphics[viewport = 0 0 260 235,height=1.9cm]{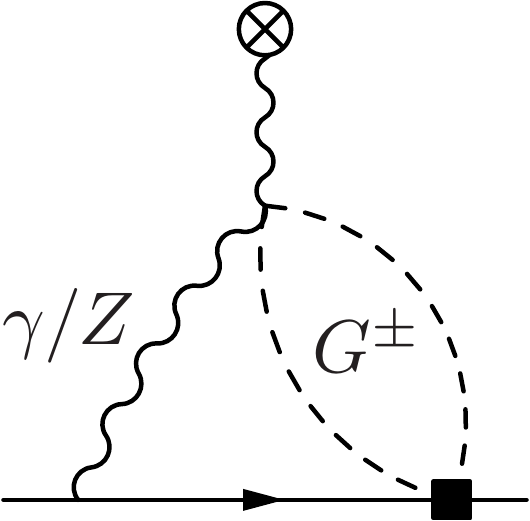}\enspace\enspace
\includegraphics[viewport = 0 0 260 235,height=1.9cm]{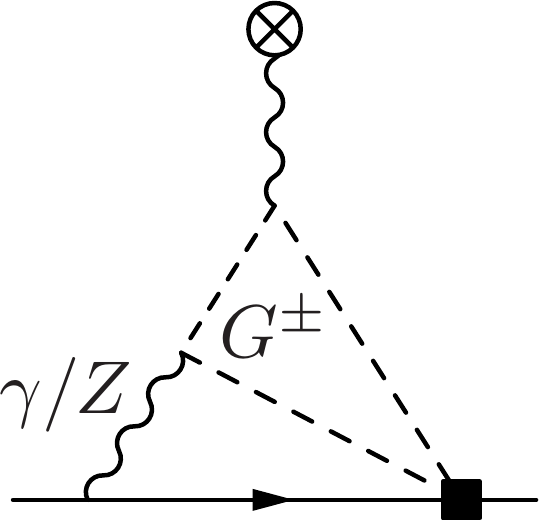}\\[5mm]
\includegraphics[viewport = 0 0 260 235,height=1.9cm]{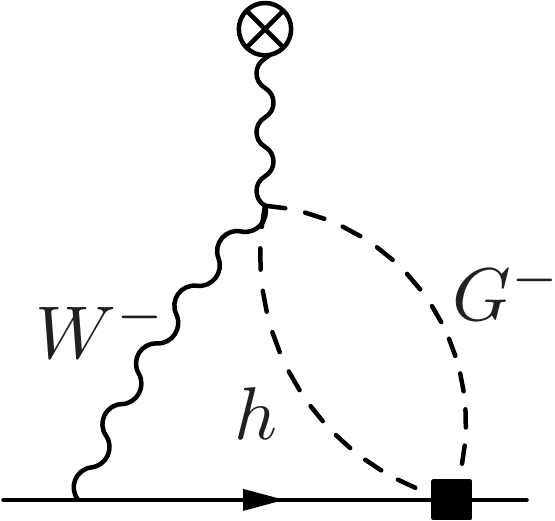}\enspace\enspace
\includegraphics[viewport = 0 0 260 235,height=1.9cm]{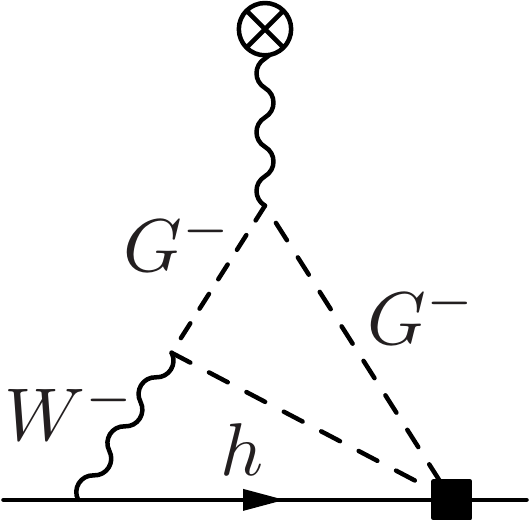}\enspace\enspace
\includegraphics[viewport = 0 0 260 235,height=1.9cm]{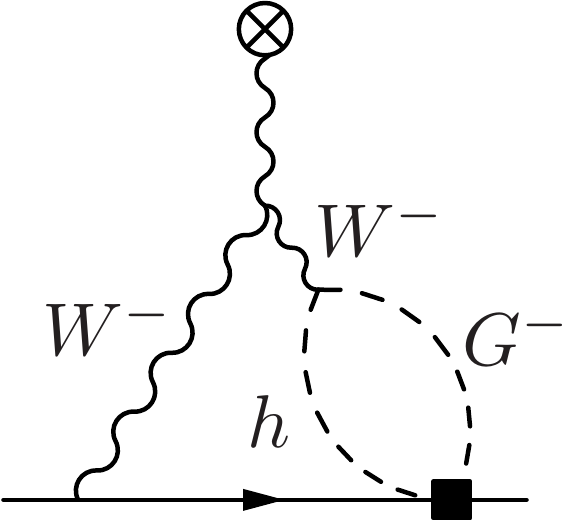}
\caption{
Diagrams involving the four-point interactions in (\ref{eq:fourptEFT}) that contain the leading logarithmic contribution to the electron EDM.}
\label{fig:Dim5Diagrams}
\end{figure}

Gauge-independence is achieved when we include the second class of interactions generated by (\ref{eq:dim6}) in the electroweak vacuum.  These are the four-point interactions involving the charged Goldstone bosons obtained by setting just one Higgs field to its vacuum expectation value
\begin{multline}\label{eq:fourptEFT}
\mathcal{L}_6 \supset -\frac{m_e}{M^2} c_\ell \sin\beta\cos\beta\,\text{Im}(\lambda_5)\times\\
\Big[ iG^+G^-\,\bar{e}\gamma_5 e
+ \big(i \sqrt{2} h G^-\,\bar{e} P_L \nu + \text{c.c.}\big) \Big]\,.
\end{multline}
These interactions generate new diagrams shown in Fig.~\ref{fig:Dim5Diagrams} and are essential to obtain a gauge-independent result.  Furthermore, we find that they are UV divergent as expected from power counting
\begin{multline}
\delta_e = \frac{v^2}{M^2} c_\ell \sin\beta\cos\beta\,\text{Im}(\lambda_5)
\Big[\frac{-3}{4 \cw^2} \Big(\frac{1}{2\epsilon} -\gamma_E + \ln(4\pi) \\ + \ln\big(\frac{\mu^2}{m_W^2}\big)+\frac{7}{4}\Big) +\big(\text{\begin{tabular}{cc}gauge dep.\\[-1mm]non-log.\end{tabular}}\big)\Big]\,,
\end{multline}
where the gauge-dependent non-logarithmic terms have been omitted for brevity.  The appearance of a simple $1/\epsilon$ pole signals the two loop mixing of the dimension-6 operator in (\ref{eq:dim6}) into the electron dipole moment operator.  This mixing effect was noted in \cite{Panico:2018hal} based on a model-independent systematic analysis of CP-violating dimension-6 operators, and the logarithm found there agrees with our explicit calculation in the C2HDM.

Our final result of the EFT calculation in dimensional regularization is the sum of both classes of diagrams, which we identify as the IR part of (\ref{eq:UVIR}) given by (\ref{eq:irFP})--(\ref{eq:IRW}).  The appearance of the dimensional regularization parameters and regularization-dependent constants in (\ref{eq:IRW}) are understood to arise from the separation into the short distance and long distance contributions based on the EFT computation just outlined.  The low energy constant associated with the electron EDM operator in the 2HDM is then given by short distance contributions $\Delta^\text{UV}_{W}+\Delta^\text{UV}_{H^+}$ in (\ref{eq:UVW}), (\ref{eq:UVH}), and serves as the counterterm for the EFT computation.  With respect to the full C2HDM calculation, it is interesting to note that the bulk of the non-logarithmic contributions are captured in the infrared by the EFT.  The only contributions that are not reproduced are those arising from the numerically small charged Higgs Barr-Zee diagrams in (\ref{eq:UVH}), and regulator-dependent constants in the $W$ loop contributions in (\ref{eq:UVW}).

Despite its complicated appearance, the electron EDM near the decoupling limit (\ref{eq:UVIR}) depends straightforwardly on a few C2HDM parameters allowing us to provide simple numerical expressions by inserting the known values of the SM parameters (\ref{eq:smParams}):
\begin{widetext}
\begin{align}
 \label{eq:decouplingType1}
 \text{Type~I:} &&  d_e &= -1.06 \times 10^{-27} e\,\text{cm} \times \left(\frac{1\,\text{TeV}}{M}\right)^2  \text{Im}(\lambda_5) \phantom{\Big\{} \cos^2\!\beta \Big[ 1 + 0.07 \ln\left(\frac{M}{1\,\text{TeV}} \right) \Big] \,, \\
 \label{eq:decouplingType2}
 \text{Type~II:} &&  d_e &= \phantom{-}0.47 \times 10^{-27} e\,\text{cm} \times \left(\frac{1\,\text{TeV}}{M}\right)^2  \text{Im}(\lambda_5) \Big\{ \sin^2\!\beta \Big[ 1 + 0.16 \ln\left(\frac{M}{1\,\text{TeV}} \right) \Big] - 1.26 \cos^2\!\beta \Big\} \,,
 \\
 \text{Lepton Specific:} && d_e &= \phantom{-}0.47 \times 10^{-27} e\,\text{cm} \times \left(\frac{1\,\text{TeV}}{M}\right)^2  \text{Im}(\lambda_5) \Big\{ \sin^2\!\beta  \Big[ 1 + 0.16 \ln\left(\frac{M}{1\,\text{TeV}} \right) \Big] - 1.25 \cos^2\!\beta \Big\} \,,
 \\
 \label{eq:decouplingFlipped}
 \text{Flipped:} &&  d_e &= -1.06 \times 10^{-27} e\,\text{cm} \times \left(\frac{1\,\text{TeV}}{M}\right)^2  \text{Im}(\lambda_5) \Big\{ \cos^2\!\beta \Big[ 1 + 0.07 \ln\left(\frac{M}{1\,\text{TeV}} \right) \Big] + 0.002 \sin^2\!\beta\Big\} \,.
\end{align}
\end{widetext}
The leading logarithmic contribution is suppressed by a small coefficient, requiring $M$ to be orders of magnitude above the TeV scale before it can dominate the nonlogarithmic contributions.  The above expressions also reveal a numerical cancellation near $\tan\beta \approx 1$ for Type~II and the Lepton Specific models, which is evident in the right panel of Fig.~\ref{fig:type1-tanbeta-plot}.

We pause to comment on a similar EFT analysis that was recently carried out in \cite{Egana-Ugrinovic:2018fpy}.  Their results differ from ours due to the omission of the diagrams of Fig.~\ref{fig:Dim5Diagrams} derived from the interactions in (\ref{eq:fourptEFT}). Consequently, their results are gauge-dependent and their formulae for the electron EDM miss the leading logarithmic contribution. The numerical effect is at the level of $\sim 25\%$ for Type~I and $\sim 55\%$ for Type~II at $\mHp \approx 1\text{ TeV}$.

\begin{figure}[tb]\centering
\includegraphics[width=0.8\columnwidth]{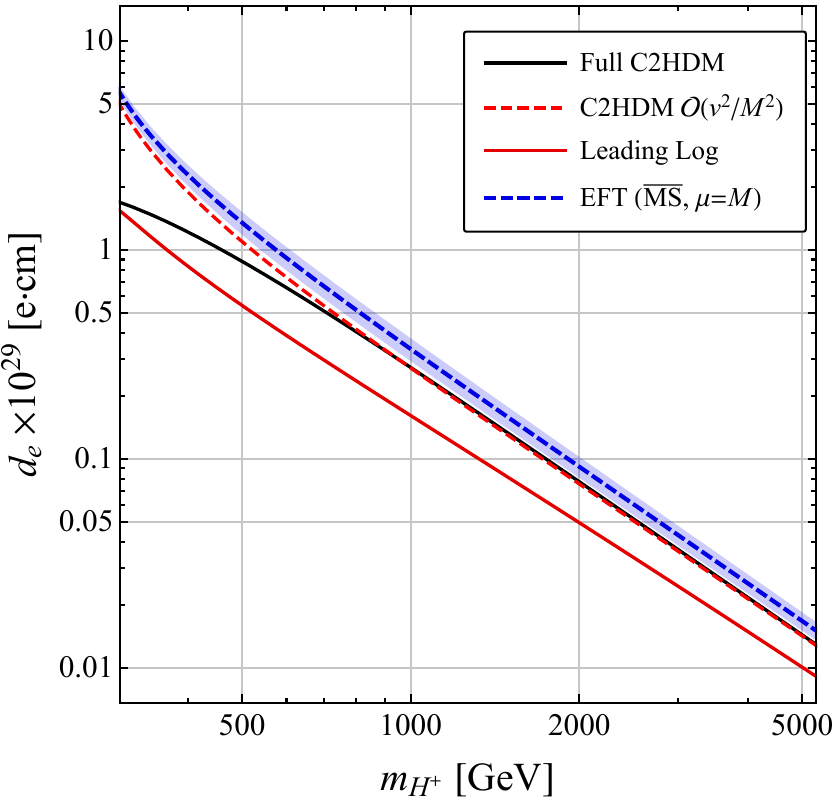}
\caption{Approximations to predictions of the electron EDM in the Type~II C2HDM as a function of $m_{H^+}$, at the benchmark point (\ref{eq:c2hdmParams}) with $\tan\beta=2$.  The black line is the full two loop result in the C2HDM (\ref{eq:gaugeinv-totaledm}).  The dashed red line is its asymptotic approximation near the decoupling limit through $\mathcal{O}(v^2/M^2)$ given in (\ref{eq:decouplingType2}). The solid red curve is the leading logarithmic approximation in (\ref{eq:leadinglogapprox}) and the dashed blue curve is the EFT result in the $\overline{\text{MS}}$ scheme given by the IR part of (\ref{eq:UVIR}) with $\mu=M$. The shaded blue region is obtained varying the scale between $\mu=M/2$ and $\mu=2M$.}
\label{fig:decoupling}
\end{figure}

In Fig.~\ref{fig:decoupling} we numerically compare various approximations to the electron EDM as a function of $\mHp$ for the Type~II C2HDM.  All other parameters are fixed according to the benchmark point in (\ref{eq:c2hdmParams}) with $\tan\beta=2$.  The black line shows the result of the full two loop calculation (\ref{eq:gaugeinv-totaledm}).  Its approximation near the decoupling limit (\ref{eq:decouplingType2}) is shown in dashed red, and asymptotically approaches the full result (black curve) as $\mHp\rightarrow\infty$.  The solid red curve shows the leading logarithmic approximation (\ref{eq:leadinglogapprox}), and for the modest values of $\mHp$ displayed in the plot, only provides the correct order of magnitude for the electron EDM.  Its approach to the black curve is slow, and good agreement is not reached until $\mHp$ is several orders of magnitude above the electroweak scale.  Finally, the EFT result in the $\overline{\text{MS}}$ scheme given by the IR part of (\ref{eq:UVIR}) with $\mu=M$ is shown in blue, with the shaded band obtained by varying the scale between $\mu=M/2$ and $\mu=2M$.  Because of its inability to capture the model-dependent non-logarithmic contributions in the UV, its approach to black curve is as slow as the leading logarithmic approximation (solid red).  However, its difference relative to the full two loop calculation is smaller since it accounts for a significant part of the non-logarithmic contributions in the IR.  

Before finishing this section, we would like to stress the limitation of the ``$\kappa$ framework'' often used in the literature to parametrize the possible effects of a CP violating SM Higgs boson on the EDMs~\cite{Harnik:2012pb,Altmannshofer:2015qra,Brod:2018lbf}.  As explained below (\ref{eq:EFT-relevant}), a modified Higgs coupling of the form $-\kappa h \bar{e} i\gamma_5 e$ by itself leads to gauge-dependent contributions to the EDM and needs to be supplemented by additional interactions of the form in (\ref{eq:fourptEFT}). However, the full gauge-independent result for the EDM that takes into account the additional interactions is found to be logarithmically divergent. The finite part of the necessary counterterm is scheme dependent and any analysis of the EDM in the EFT framework beyond the leading logarithms is therefore model dependent.

\section{Summary}\label{Sec:summary}
In this paper, we presented the first complete two loop calculation of the electron EDM in the complex two-Higgs doublet model.
We calculated the EDM in two separate classes of gauge, and obtained identical gauge-independent results.  Our final formula is given in (\ref{eq:gaugeinv-totaledm}) which we reproduce here for reference
\begin{multline}
\frac{d_e}{e} = \frac{\sqrt{2}\alpha G_F m_e}{64 \pi^3}\times\\
\big[\sum_f(\delta^\text{EM}_f+\delta^\text{NC}_f) + (\delta_{H^+}^\text{EM}
+ \delta_{H^+}^\text{NC} + \delta_{H^+}^\text{CC})\\
+ (\delta_{W}^\text{EM} + \delta_{W}^\text{NC}  +
\delta_{W}^\text{CC}  + \delta_\text{kite}^\text{NC} + \delta_\text{kite}^\text{CC} )\big]\,,
\end{multline}
The individual contributions are given in (\ref{eq:emBZ-fermion}), (\ref{eq:ncBZ-fermion}), (\ref{eq:emBZ-chargedHiggs}), (\ref{eq:ncBZ-chargedHiggs}), (\ref{eq:ccBZ-chargedHiggs}), (\ref{eq:emBZ-W_2}), (\ref{eq:ncBZ-W_2}), (\ref{eq:ccBZ-W_2}), (\ref{eq:ncKite}), and (\ref{eq:ccKite}). We collect these expressions in a \emph{Mathematica} notebook that is provided as ancillary material.

Compared with the most recent evaluation of the electron EDM by Abe \emph{et. al.} \cite{Abe:2013qla}, our calculation incorporated the kite contributions in Fig.~\ref{fig:kiteDiagrams}. Generically, these new contributions lead to $\mathcal O(1)$ corrections to the prediction of the electron EDM (see for example Fig. \ref{fig:type1-tanbeta-plot}), and they are particularly relevant in the Type~II and Lepton Specific CHDMs.
 In the Type~II and Lepton Specific C2HDMs there are regions in parameter space where the fermion and gauge loop contributions interfere destructively causing the electron EDM to dip below current limits established by the ACME collaboration.  We found that the inclusion of the kite diagrams can significantly shift the location of these cancellations.

In addition to the full result, we derived the leading order asymptotic expansion of the electron EDM near the decoupling limit.  The expressions for common types of C2HDMs are provided in Eqs.~(\ref{eq:decouplingType1})--(\ref{eq:decouplingFlipped}).  We find that the electron EDM exhibits a logarithmic dependence on the heavy masses.  From the point of view of an EFT, the logarithm indicates sensitivity to the UV scale implying that the precise prediction of the EDM cannot be determined in a model independent manner.  However, for the case of the C2HDM we find that a large part of the electron EDM near the decoupling limit is reproduced in the infrared.

Furthermore, we have emphasized that the analysis of the electron EDM based on a simple phenomenological parameterization of CP-violating electron Yukawa coupling $-\kappa h \bar{e} i\gamma_5 e$ requires caution since the resulting prediction of the electron EDM is not gauge-invariant.

As explained in Sec.~\ref{Sec:Lightquark}, the formulae for the electron EDM are easily adaptable for EDMs of light quarks.  It would be interesting to have a calculation of EDMs for the heavier bottom and top quarks, which require separate treatment.  Also, it would be interesting to perform a full calculation of the electron EDM for other types of 2HDMs without a softly broken $\mathbb{Z}_2$ symmetry, or in which CP is spontaneously broken.  We leave these exercises to future work.

\acknowledgments
We thank Joachim Brod for helpful discussions.  We thank author Teppei Kitahara of Abe \emph{et. al} \cite{Abe:2013qla} for alerting us that our result for $\delta_W^\text{NC}(\xi=1)$ does agree with theirs, contrary to our claim in a previous version of our paper.
We thank the authors of~\cite{Davila:2025goc} for pointing out a typo in equation~\eqref{eq:dim6} in previous versions of our paper.

The research of WA was partly supported by the National Science Foundation under Grant No. NSF 1912719; the research of SG is supported in part by the National Science Foundation CAREER grant PHY-1915852; the research of NH was supported in part by the National Science Foundation Grant No. NSF 1912719 and the CAREER grant PHY-1915852; and the research of HHP was supported by the U.S. Department of Energy grant number DE-FG02-04ER41286 and by the National Science Foundation Grant No. NSF 1912719.

\setcounter{secnumdepth}{0}
\appendix
\section{Appendix: Parameters of the Higgs Potential}

In this appendix, we collect useful equations on the 2HDM scalar potential  \cite{Davidson:2005cw}. First, the conditions of minimization of the potential in (\ref{eq:genpot})
\begin{eqnarray}\nonumber
m_{11}^2 &=& {\rm{Re}}(m_{12}^2\,e^{i\zeta})\,\frac{v_2}{v_1}  -\frac{1}{2}
\left[\lambda_1v_1^2+\lambda_{345}\,v_2^2\right]\,,
\label{minconditionsa} \\[6pt]\nonumber
m_{22}^2&=& {\rm{Re}}(m_{12}^2\,e^{i\zeta})\,\frac{v_1}{v_2} -\frac{1}{2}
\left[\lambda_2 v_2^2+\lambda_{345}\,v_1^2
\right],\\
&&{\rm{Im}}(m_{12}^2 e^{i\zeta})=\frac{v_1v_2}{2}{\rm{Im}}(\lambda_5 e^{2i\zeta})\,,
\label{minconditionsb}
\end{eqnarray}
can be used to determine $v_1$, $v_2$ and $\zeta$, where $\lambda_{345} = \lambda_3 + \lambda_4 + \text{Re}(\lambda_5 e^{2i\zeta})$.  Utilizing these minimization conditions, we note that the C2HDM Higgs potential is fully determined by 9 independent free parameters, for example by the set $\tan\beta,{\rm{Re}}(m_{12}^2),\lambda_1,\lambda_2,\lambda_3,\lambda_4,{\rm{Re}}(\lambda_5),{\rm{Im}}(\lambda_5), v(=246$ GeV).

The Higgs potential can also be expressed in the Higgs basis defined in (\ref{eq:HBpot}). The corresponding mass terms and quartic interactions are linearly related to the $\lambda_i$, $m_{ij}^2$:
\begin{equation}
\begin{aligned}
Y_1={}&m_{11}^2 c_\beta^2 +m_{22}^2 s_\beta^2-{\rm{Re}}(m_{12}^2 e^{i\zeta})s_{2\beta}\\
Y_2={}&m_{11}^2 s_\beta^2 +m_{22}^2 c_\beta^2+{\rm{Re}}(m_{12}^2 e^{i\zeta})s_{2\beta}\\
Y_3 e^{i\zeta}={}&\textstyle\frac{1}{2}(m_{11}^2-m_{22}^2)s_{2\beta}+{\rm{Re}}(m_{12}^2 e^{i\zeta})c_{2\beta}\\
 &\hspace{3cm}
+i\,{\rm{Im}}(m_{12}^2 e^{i\zeta})
\end{aligned}
\end{equation}
\begin{gather}
\begin{aligned}
    \label{eq:Zlist}
Z_1 ={}& \lambda_1 c^4_\beta +\lambda_2 s^4_\beta +\textstyle\frac{1}{2}\lambda_{345}s_{2\beta}^2 \\
Z_2 ={}& \lambda_1s^4_\beta +\lambda_2 c^4_\beta + \textstyle\frac{1}{2}\lambda_{345}s_{2\beta}^2\\
Z_3 ={}&\textstyle \frac{1}{4}s_{2\beta}^2\left(\lambda_1+\lambda_2-2\lambda_{345}\right)+\lambda_3\\
Z_4 ={}&\textstyle \frac{1}{4}s_{2\beta}^2\left(\lambda_1+\lambda_2-2\lambda_{345}\right)+\lambda_4 \\
Z_5 e^{2i\zeta}={}&\textstyle \frac{1}{4}s_{2\beta}^2\left(\lambda_1+\lambda_2-2\lambda_{345}\right) \\
&\hspace{15mm}+{\rm{Re}}(\lambda_5 e^{2i\zeta}) +i c_{2\beta}{\rm{Im}}(\lambda_5 e^{2i\zeta})
\end{aligned}\\
\begin{aligned}
\nonumber Z_6 e^{i\zeta}={}& -\textstyle\frac{1}{2}s_{2\beta}\big(\lambda_1 c_\beta^2 -\lambda_2 s_\beta^2-\lambda_{345} c_{2\beta} -i{\rm{Im}}(\lambda_5 e^{2i\zeta})\big)\\
\nonumber Z_7 e^{i\zeta}={}& -\textstyle\frac{1}{2} s_{2\beta}\big(\lambda_1 s_\beta^2 -\lambda_2 c_\beta^2 +\lambda_{345} c_{2\beta}+i{\rm{Im}}(\lambda_5 e^{2i\zeta})\big).
\end{aligned}
\end{gather}
The set of 9 independent parameters that we choose for our numerics that determine the Higgs potential is given by: $\tan\beta,~m_{H^+}$, $\text{Im}(\lambda_5 e^{2i\zeta})$, $Z_3$, $Z_4$, $\text{Re}(Z_5 e^{2i\zeta})$, $\text{Re}(Z_6 e^{i\zeta})$, with $\zeta=0$, after fixing $m_1$ and $v$. For completeness, we provide the remainder of the $Z_i$ in terms of our chosen set, having set $\zeta = 0$. From the last three equations of (\ref{eq:Zlist}), $\rm{Im} (Z_{5,6,7})$ are determined. Utilizing results in \cite{Boto:2020wyf}, with a deriviation given in the supplementary \emph{Mathematica} notebook, the remaining quartic couplings are given by

    \begin{align}
        \nonumber
        \rm{Re}(Z_7) &= \rm{Re}(Z_6) + \frac{\rm{Im}(Z_5) \rm{Im} \! \left[Z_6^* (Z_1 -Z_3 -Z_4 - Z_5)\right]}{ 2\, \rm{Im}(Z_6)^2} \, ,\\
        Z_2 &= Z_1 + (2 / t_{2\beta}) \left(Z_6 + Z_7 \right)\,,
    \end{align}
where $Z_1$ is determined diagonalizing the mass matrix in (\ref{eq:neutralMassMtx}) and imposing $m_h=125$ GeV. 

\bibliography{references}{}

\end{document}